\documentstyle[prd,eqsecnum,aps,amssymb]{revtex}
\def\be{\begin{equation}}
\def\ee{\end{equation}}
\def\bea{\begin{eqnarray}}
\def\eea{\end{eqnarray}}
\def\l{\label}
\def\ct{\cite}
\def\r{\ref}
\def\bi{\bibitem}
\def\lgth{[\,\mbox{length}\,]}
\def\gam{\gamma}
\def\d{\delta}
\def\eps{\epsilon}
\def\gam{\gamma}
\def\Gam{\Gamma}
\def\Th{\Theta}
\def\sig{\sigma}
\def\sigp{\sigma_{+}}
\def\sigm{\sigma_{-}}
\def\sigc{\sigma_{\times}}
\def\sigtw{\sigma_{2}}
\def\sigth{\sigma_{3}}
\def\np{n_{+}}
\def\nm{n_{-}}
\def\nc{n_{\times}}
\def\ntw{n_{2}}
\def\nth{n_{3}}
\def\om{\omega}
\def\Om{\Omega}
\def\udot{\dot{u}}
\def\cs{c_{s}}

\def\p{{\bf e}}
\def\ptl{\partial}
\def\la{\langle}
\def\ra{\rangle}

\def\hsp5{\hspace{5mm}}
\newcommand{\sfrac}[2]{{\textstyle{#1\over#2}}}
\def\case#1/#2{\textstyle\frac{#1}{#2}}
\def\cqg{{Class. Quantum Grav.} }
\def\grg{{Gen. Rel. Grav.} }
\def\prd{{Phys. Rev. D} }
\def\prl{{Phys. Rev. Lett.} }
\def\jmp{{J. Math. Phys.} }
\newcommand{\enl}{\\\hfill\rule{0pt}{0pt}}
\begin{document}
\preprint{uct--cosmology--00/06, AEI--2000--039}
\draft
%%%%%%%%%%%%%%%%%%%%%%%%%%%%%%%%%%%%%%%%%%%%%%%%%%%%%%%%%%%%%%%%%%%
\title{On the propagation of jump discontinuities in relativistic
cosmology}
%%%%%%%%%%%%%%%%%%%%%%%%%%%%%%%%%%%%%%%%%%%%%%%%%%%%%%%%%%%%%%%%%%%
\author{Henk van Elst,${}^{1}${}\thanks{Electronic address:
henk@gmunu.mth.uct.ac.za}\,
George F. R. Ellis,${}^{1}${}\thanks{Electronic address:
ellis@maths.uct.ac.za}\,
and Bernd G. Schmidt${}^{2}${}\thanks{Electronic address:
bernd@aei-potsdam.mpg.de}}

\address{${}^{1}\!$Cosmology Group, Department of Mathematics and
Applied Mathematics, University of Cape Town\\ Rondebosch 7701,
Cape Town, South Africa}

\address{${}^{2}\!$Albert--Einstein--Institut,
Max--Planck--Institut f\"{u}r Gravitationsphysik, Am M\"{u}hlenberg
1\\14476 Golm, Germany}

%%%%%%%%%%%%%%%%%%%%%%%%%%%%%%%%%%%%%%%%%%%%%%%%%%%%%%%%%%%%%%%%%%%
\date{July 3, 2000}
%%%%%%%%%%%%%%%%%%%%%%%%%%%%%%%%%%%%%%%%%%%%%%%%%%%%%%%%%%%%%%%%%%%
\maketitle
%%%%%%%%%%%%%%%%%%%%%%%%%%%%%%%%%%%%%%%%%%%%%%%%%%%%%%%%%%%%%%%%%%%
%%%%%%%%%%%%%%%%%%%%%%%%%%%%%%%%%%%%%%%%%%%%%%%%%%%%%%%%%%%%%%%%%%%
\begin{abstract}
%%%%%%%%%%%%%%%%%%%%%%%%%%%%%%%%%%%%%%%%%%%%%%%%%%%%%%%%%%%%%%%%%%%
A recent dynamical formulation at derivative level $\ptl^{3}g$ for
fluid spacetime geometries $\left(\,{\cal M},\,{\bf g},\,{\bf
u}\,\right)$, that employs the concept of evolution systems in
first-order symmetric hyperbolic format, implies the existence in
the Weyl curvature branch of a set of timelike characteristic
3-surfaces associated with propagation speed $|\,v\,| =
\sfrac{1}{2}$ relative to fluid-comoving observers. We show it is
the physical role of the constraint equations to prevent
realisation of jump discontinuities in the derivatives of the
related initial data so that Weyl curvature modes propagating along
these 3-surfaces cannot be activated. In addition we introduce a
new, illustrative first-order symmetric hyperbolic evolution system
at derivative level $\ptl^{2}g$ for baryotropic perfect fluid
cosmological models that are invariant under the transformations of
an Abelian $G_{2}$ isometry group.

%%%%%%%%%%%%%%%%%%%%%%%%%%%%%%%%%%%%%%%%%%%%%%%%%%%%%%%%%%%%%%%%%%%
\end{abstract}
%%%%%%%%%%%%%%%%%%%%%%%%%%%%%%%%%%%%%%%%%%%%%%%%%%%%%%%%%%%%%%%%%%%
%\narrowtext
\widetext
%%%%%%%%%%%%%%%%%%%%%%%%%%%%%%%%%%%%%%%%%%%%%%%%%%%%%%%%%%%%%%%%%%%
\section{Introduction}
\l{sec:intro}
%%%%%%%%%%%%%%%%%%%%%%%%%%%%%%%%%%%%%%%%%%%%%%%%%%%%%%%%%%%%%%%%%%%
It is well known that the mathematical nature of the evolution
system within the relativistic gravitational field equations is
essentially {\em hyperbolic\/}, with domains of dependence and
influence determined by the speed of light. However, this nature is
not obvious when the dynamical equations are written out in
standard form, employing either a metric approach
\ct{mtw73}, a Hamiltonian (ADM) representation \ct{adm62}, a
$1+3$ orthonormal frame (ONF) formulation \ct{ell67,mac73}, or a
form obtained from a covariant $(1+3)$--decomposition
\ct{ehl61,ell71}. Consequently, over the years considerable effort
has gone into determining ways of making this hyperbolic nature
clear. Indeed, evolution systems of partial differential equations
in {\em first-order symmetric hyperbolic\/} (FOSH) format have
proven to be a main theme of research for at least the last five
years. This activity was particularly motivated by the desire to
carry over to the numerical investigation of relativistic effects
such as generation of gravitational radiation associated with the
in-spiralling of black hole and neutron star binaries the methods
and expertise gained in areas of computational physics with a
longer tradition like, e.g., hydrodynamics, where evolution systems
of FOSH format are commonplace (see, e.g., the recent reviews
\ct{reu98} and \ct{friren00}).

One of the most promising methods for obtaining a FOSH
representation for the relativistic gravitational field equations
is to use an extended $1+3$ orthonormal frame formulation that
includes the once-contracted second Bianchi identities
\ct{hveugg97}. In terms of the highest-order partial derivatives of
the spacetime metric ${\bf g}$ implicitly occurring, the dynamical
equations here rank at derivative level $\ptl^{3}g$. The set of
geometrically defined field variables contains the components of
the physically significant Weyl curvature tensor, and so we refer
to this as a $1+3$ orthonormal frame {\em curvature\/} formulation
of gravitational fields. It represents a completion of the
physically and geometrically transparent $1+3$ covariant formalism
for the cosmological case, in which all field variables are
covariantly defined relative to a uniquely defined future-directed
timelike reference congruence given by the unit 4-velocity field
${\bf u}$ of matter present \ct{ell71,ellhve99}. As Friedrich first
showed, when the matter source for a spacetime geometry
$\left(\,{\cal M},\,{\bf g},\,{\bf u}\,\right)$ is described
phenomenologically as a perfect fluid, or one restricts to vacuum
situations, upon introduction of a set of local coordinates and a
specific choice to remove the gauge fixing freedom it is indeed
possible to find linear combinations of the field variables and
their dynamical equations that lead to an evolution system of
(autonomous) partial differential equations in FOSH format, as
desired \ct{fri96,fri98}. It is puzzling, then, that when this is
done, as well as the expected sets of characteristic 3-surfaces
associated with propagation speeds relative to ${\bf u}$ of
$|\,v\,| = 0$ (Coulomb-like gravitational and fluid rotational
modes), $|\,v\,| = \cs$ (sound wave modes), and $|\,v\,| = 1$
(transverse gravitational wave modes), the equations on the face of
it indicate that the semi-longitudinal Weyl curvature
characteristic eigenfields, identified long ago by Szekeres
\ct{sze65}, appear related to an additional set of timelike
characteristic 3-surfaces associated with propagation speed
relative to ${\bf u}$ of one-half the speed of light, $|\,v\,| =
\sfrac{1}{2}$.

Friedrich assumed that these modes could be ignored, because, at
least in the {\em vacuum case\/}, they had no invariant meaning
\ct{fri96}. Two of us (HvE and GFRE), on the other hand, argued in
a recent paper that in the fluid context that is relevant in
relativistic cosmology, this is not the case, essentially because
(i) the timelike reference congruence corresponding to ${\bf u}$ is
invariantly defined by the presence of matter, and (ii) the
hypothetical Weyl curvature modes cannot all be transformed to zero
by using up the remaining gauge fixing freedom \ct{hveell99}. Thus,
these might possibly represent propagation modes that are
physically meaningful. In private communications, Friedrich
responded that the form of the $\ptl^{3}g$-order FOSH evolution
systems discussed in Refs. \ct{fri98} and \ct{hveell99} is {\em
not\/} unique; one could obtain other reductions of the
relativistic gravitational field equations that would lead to
additional timelike characteristic 3-surfaces associated with
propagation speeds {\em different\/} from $|\,v\,| =
\sfrac{1}{2}$. And, hence, these modes cannot be physical (see also
Ref. \ct{friren00}).  However, that argument is not very
satisfactory. In our view, it compounds the problem, rather than
solving it, by suggesting (if we take for face value the FOSH
formalism and its interpretation) that still other (subluminal)
propagation speeds could be associated with the transport of
physical quantities, as well as $|\,v\,| =
\sfrac{1}{2}$. The implication is that any single FOSH evolution
system cannot be taken seriously without considering the set of
{\em all possible\/} FOSH evolution systems, which is difficult to
determine.

What had not been fully taken into account in these discussions,
even though Sec. III D 2 of Ref. \ct{hveell99} made a brief
reference to this aspect, is the physical role the {\em constraint
equations\/} play in the selection of appropriate classes of
initial data sets. Specifically, what kind of physically relevant
{\em jump discontinuities\/} they do allow to be present in the
derivatives of these data sets \ct{pir57}. Jump discontinuities in
the derivatives of the initial data are of major dynamical
importance as they can be thought of as representing the triggering
or terminating events of physical generation
processes.\footnote{J\"{u}rgen Ehlers (private communication).} In
the present paper we show that an analysis of the constraint
equations leads to a resolution of the problem: {\em while the
additional timelike characteristic 3-surfaces do indeed occur in
the $\ptl^{3}g$-order FOSH evolution system, the constraint
equations are of such a form as to prevent the corresponding
semi-longitudinal Weyl curvature modes being activated\/}. Because
jump discontinuities {\em cannot\/} exist in the values of the
constraint equations themselves (they have to be zero everywhere),
jump discontinuities in the derivatives of the initial data can
only be propagated along the characteristic 3-surfaces associated
with propagation speeds relative to ${\bf u}$ of $|\,v\,| = 0$,
$|\,v\,| = \cs$, and $|\,v\,| = 1$, but not along those associated
with $|\,v\,| = \sfrac{1}{2}$. Hence, one cannot physically send
arbitrary information along the latter set of timelike
characteristic 3-surfaces.

The problem of additional characteristic 3-surfaces and how the
constraint equations select appropriate initial data sets can be
highlighted in a nice and transparent fashion by a suggestive
example that derives from a set of linearised relativistic
gravitational field equations given in a paper by Kind, Ehlers and
Schmidt \ct{kinetal93}. We will briefly discuss this example in
Appendix \r{app1}.

A possible complementary explanation to the above viewpoint arises
from the following chain of arguments:

(i) For vacuum as well as perfect fluid spacetime geometries, there
exist $\ptl^{2}g$-order dynamical formulations of the relativistic
gravitational field equations (with harmonic coordinate gauge
fixing) in which characteristic 3-surfaces {\em only\/} correspond
to the local light cones, the local sound cones, or are generated
by the fluid flow lines. (See, e.g., Refs. \ct{reu98} and
\ct{friren00}.)

(ii) There are dynamical formulations of order $\ptl^{2}g$ or
$\ptl^{3}g$ that have {\em additional\/} (timelike or spacelike)
characteristic 3-surfaces. Along these characteristic 3-surfaces
the evolution system will propagate jump discontinuities if we put
them into the related initial data ({\em ignoring\/} the constraint
equations).

(iii) Different dynamical formulations with different sets of
characteristic 3-surfaces all describe for {\em identical\/}
initial data the {\em same\/} spacetime geometry.

(iv) {\em Because\/} of the existence of a dynamical formulation
according to (i), jump discontinuities as described in (ii) {\em
cannot\/} really exists (and so be physically meaningful). Hence,
the only logical possibility is that in (ii) the constraint
equations prevent the existence of such jump discontinuities.

The present paper shows that the semi-longitudinal Weyl curvature
modes that potentially exist cannot in fact carry ``gravitational
news''. Of course, this is expected, so this analysis simply
confirms what everyone has believed all along. However, we believe
it is nevertheless useful, both in terms of showing important
relations between the constraint equations and the set of
characteristic 3-surfaces that can occur in general FOSH evolution
systems (and have not been clearly demonstrated in the literature),
and because this does indeed resolve the issue at hand in the
specific important case of relativistic gravitation, where the
variables presented here, and hence the associated dynamical
equations, have much to recommend them. Without the analysis given
in this paper, the apparent ($|\,v\,| = \sfrac{1}{2}$)--modes
remain an annoying and unresolved problem.

The outline of the paper is as follows. In Sec. \r{sec:maths} we
briefly discuss the mathematical concepts relevant to our analysis:
we review the central ideas behind FOSH evolution systems, we
address the issue of gauge fixing freedom in a $1+3$ orthonormal
frame formulation of the relativistic gravitational field
equations, we list the set of constraint equations we need to
consider, and we introduce a $(1+1+2)$--decomposition of all
geometrically defined field variables. Then, in Sec. \r{sec:jumps},
we address from a purely {\em local\/} viewpoint the question of
how the constraint equations select appropriate initial data sets
for the $\ptl^{3}g$-order FOSH evolution system introduced in
Ref. \ct{hveell99}. To the best of our knowledge, we present in
Sec. \r{sec:g2fosh} for the first time a fully gauge-fixed
autonomous FOSH evolution system for a special class of spatially
inhomogeneous perfect fluid cosmological models on the basis of a
$1+3$ orthonormal frame formulation that is only of order
$\ptl^{2}g$ in the degrees of freedom of the gravitational
field.\footnote{The equations of motion for the matter sources will
necessarily derive from the twice-contracted second Bianchi
identities and so are typically of order $\ptl^{3}g$.} We refer to
this as a $1+3$ orthonormal frame {\em connection\/} formulation of
gravitational fields. Additionally, we derive for these
cosmological models, for which there exists an Abelian $G_{2}$
isometry group, the transport equations that describe how
physically relevant jump discontinuities in the derivatives of the
initial data are propagated along the bicharacteristic rays of the
setting. Our conclusions are contained in Sec. \r{sec:conc}.
Finally, besides the above-mentioned suggestive example in Appendix
\r{app1}, we give in Appendix \r{app2} in explicit form those
linear combinations of the components of the extended $1+3$
orthonormal frame constraint equations that prove suitable for our
analysis.

We will use the same conventions, units and notations as introduced
in Appendix A 1 of Ref. \ct{hveell99}. \enl

%%%%%%%%%%%%%%%%%%%%%%%%%%%%%%%%%%%%%%%%%%%%%%%%%%%%%%%%%%%%%%%%%%%
\section{Mathematical preliminaries}
\l{sec:maths}
%%%%%%%%%%%%%%%%%%%%%%%%%%%%%%%%%%%%%%%%%%%%%%%%%%%%%%%%%%%%%%%%%%%
%------------------------------------------------------------------
\subsection{First-order symmetric hyperbolic evolution systems}
\l{subsec:fosh}
%------------------------------------------------------------------
We consider evolution systems for a collection of $k$ real-valued
field variables $u^{A} = u^{A}(x^{\mu})$ that are composed of a set
of $k$ quasi-linear partial differential equations of {\em first
order\/} given by
\be
\l{fosh}
M^{AB\,\mu}(x^{\nu},u^{C})\,\ptl_{\mu}u_{B} = N^{A}(x^{\nu},u^{C})
\ , \hsp5 A, \,B, \,C = 1, \dots, k \ ;
\ee
the field variables $u^{A}$ are functions of a set of local
spacetime coordinates $\{\,x^{\mu}\,\}$. Evolution systems of this
form are called {\em symmetric\/} if the real-valued $k \times k$
coefficient matrices entering the principle part satisfy
$M^{AB\,\mu} = M^{(AB)\,\mu}$; moreover, they are called {\em
hyperbolic\/} if the contraction $M^{AB\,\mu}\,n_{\mu}$ with the
coordinate components of an arbitrary {\em past-directed\/}
timelike 1-form $n_{a}$ yields a positive-definite matrix. We
remark that (i) cases with $M^{AB\,\mu} = M^{AB\,\mu}(x^{\nu})$ are
referred to as {\em semi-linear\/}, and (ii) cases with
$M^{AB\,\mu} = M^{AB\,\mu}(u^{C})$ and $N^{A} = N^{A}(u^{C})$ are
referred to as {\em autonomous\/}. In general, it proves convenient
to consider a $(1+3)$--decomposition of Eq. (\r{fosh}) in the
format
$$
M^{AB\,0}(t,x^{j},u^{C})\,\ptl_{t}u_{B}
+ M^{AB\,i}(t,x^{j},u^{C})\,\ptl_{i}u_{B}
= N^{A}(t,x^{j},u^{C}) \ .
$$
The concept of FOSH evolution systems was first introduced by
Friedrichs \ct{fri54}; a standard reference, broughtly discussing
its usefulness to applications in mathematical physics and also
presenting proofs for existence and uniqueness of solutions, is the
book by Courant and Hilbert \ct{couhil62}. The {\em characteristic
condition\/}
\be
\l{charcond}
0 = Q := \det\,[\,M^{AB\,\mu}\,\nabla_{\mu}\phi\,]
\ee
determines the coordinate components of the {\em past-directed\/}
normals $\nabla_{a}\phi$ of the set of {\em characteristic
3-surfaces\/} ${\cal C}$:$\{\phi(x^{\mu})=\mbox{const}\}$
associated with the FOSH evolution system (\r{fosh}). With
$M^{AB\,\mu} = M^{(AB)\,\mu}$, hyperbolicity of Eq. (\r{fosh}) thus
also corresponds to all individual roots (``eigenvalues'') $v$ of
Eq. (\r{charcond}) being {\em real-valued\/}. Every individual $v$
then defines a pair of so-called {\em left\/} and {\em right
nullifying vectors\/}, $l^{A}$ and $r^{A}$, by
\be
\l{lrnull}
0 =: l_{A}\,(M^{AB\,\mu}\,\nabla_{\mu}\phi) \ , \hsp5
0 =: (M^{AB\,\mu}\,\nabla_{\mu}\phi)\,r_{B} \ ;
\ee
the linearly independent sets $\{\,l^{A}\,\}$ or $\{\,r^{A}\,\}$
form a basis of the $k$-dimensional space of field variables
$u^{A}$.

According to the theory discussed in Ch. VI.4.2 of Courant and
Hilbert \ct{couhil62}, FOSH evolution systems of the format
(\r{fosh}) have the power to describe the physical transport along
so-called {\em bicharacteristic rays\/} of {\em jump
discontinuities\/} that exist in the {\em outward first
derivatives\/} across a characteristic 3-surface ${\cal
C}$:$\{\phi(x^{\mu}) =\mbox{const}\}$ of the field variables
$u^{A}$; the tangential first derivatives of the $u^{A}$ as well as
the $u^{A}$ themselves are assumed to be {\em continuous\/} across
${\cal C}$:$\{\phi(x^{\mu})=\mbox{const}\}$. As is standard, we
will use the notation
$$
[f] := \lim_{\phi\rightarrow c_{+}}f
- \lim_{\phi\rightarrow c_{-}}f = f_{+} - f_{-}
$$
to symbolise a jump discontinuity (of finite magnitude) across
${\cal C}$:$\{\phi(x^{\mu}) =\mbox{const}\}$ in the value of a
given variable $f$. Under the stated assumptions, it follows from
Eqs. (\r{fosh}) and (\r{lrnull}) that
\be
\l{jumpisr}
0 = (M^{AB\,\mu}\,\nabla_{\mu}\phi)\,[\,\ptl_{\phi}u_{B}\,]
\hsp5 \Leftrightarrow \hsp5
[\,\ptl_{\phi}u^{A}\,] = [\ptl_{\phi}u]\,r^{A} \ ,
\ee
i.e., the jump discontinuity $[\,\ptl_{\phi}u^{A}\,]$ must be
proportional to a right nullifying vector $r^{A}$. The real-valued
scalar of proportionality, denoted by $[\ptl_{\phi}u]$, is assumed
to have continuous first derivatives. Then, according to
Chs. VI.4.2 and VI.4.9 of Ref. \ct{couhil62}, for linear,
semi-linear and quasi-linear FOSH evolution systems (\r{fosh}) the
{\em transport equation\/} for $[\ptl_{\phi}u]$ along
bicharacteristic rays within the characteristic 3-surfaces ${\cal
C}$:$\{\phi(x^{\mu})=\mbox{const}\}$ takes the effective form
\be
\l{jumptrans}
0 = (l_{A}M^{AB\,\mu}r_{B})\,\ptl_{\mu}[\ptl_{\phi}u]
+ \left(\,(l_{A}M^{AB\,\mu})\,\ptl_{\mu}r_{B}
- (l_{A}N^{AB}r_{B})\,\right)\,[\ptl_{\phi}u] \ .
\ee
Note, in particular, the {\em involutive character\/} of this
relation; if $[\ptl_{\phi}u]$ is non-zero at one point along a
bicharacteristic ray, it will be non-zero everywhere along this
ray, and vice versa. Note also that the present treatment of jump
discontinuities breaks down when the ${\cal
C}$:$\{\phi(x^{\mu})=\mbox{const}\}$ within a given family start to
intersect and so prompt the formation of so-called ``shocks''.
Shock formation, however, cannot arise when the principal part of
Eq. (\r{fosh}) is semi-linear. It is a special feature of the
relativistic gravitational field equations that related FOSH
evolution systems {\em do have\/}, in the branches that evolve the
degrees of freedom in the gravitational field itself, principal
parts which are effectively semi-linear.\footnote{See, e.g.,
Eqs. (3.21) -- (3.23) in Ref. \ct{hveell99}, or Eqs. (\r{sigmdot})
-- (\r{nmdot}) in the Abelian $G_{2}$ example given in
Sec. \r{sec:g2fosh} below.} More precisely, the coefficient
matrices $M^{AB\,\mu}$ in these branches depend only on those field
variables $u^{A}$ which form a background for gravitational
dynamics in the sense of Geroch \ct{ger96}.

%------------------------------------------------------------------
\subsection{Choice of gauge source functions and local coordinates}
\l{subsec:gauge}
%------------------------------------------------------------------
As Friedrich emphasised in Sec. 5.2 of Ref. \ct{fri96}, there
exists within a $1+3$ orthonormal frame representation of the
relativistic gravitational field equations a set of ten so-called
{\em gauge source functions\/}, $G := \{\,T^{0}, \,T^{\alpha},
\,T^{\alpha}{}_{0}, \,T^{\alpha}{}_{\beta}\,\}$, that can be {\em
arbitrarily prescribed\/} in any dynamical consideration (and are
thus assumed to be ``known'').  These relate to (i) the arbitrary
choice of a {\em future-directed\/} reference ``time flow vector
field'' ${\bf T}$,\footnote{The reference vector field ${\bf T}$
need not necessarily be timelike.} which, in terms of the $1+3$ ONF
basis $\{\,\p_{0}, \,\p_{\alpha}\,\}$, is expressed by
\be
{\bf T} := T^{0}\,\p_{0} + T^{\alpha}\,\p_{\alpha} \ , \hsp5
T^{0} > 0 \ ,
\ee
and (ii) the propagation of the $1+3$ ONF basis $\{\,\p_{0},
\,\p_{\alpha}\,\}$ along ${\bf T}$, described by
\be
\nabla_{\bf T}\p_{0} := T^{\alpha}{}_{0}\,\p_{\alpha} \ , \hsp5
\nabla_{\bf T}\p_{\alpha} := T^{0}{}_{\alpha}\,\p_{0}
+ T^{\beta}{}_{\alpha}\,\p_{\beta} \ .
\ee
Parallel transport of $\{\,\p_{0}, \,\p_{\alpha}\,\}$ along ${\bf
T}$, for example, thus corresponds to setting $0 = T^{0}{}_{\alpha}
= T^{\alpha}{}_{\beta}$. Upon introduction of a dimensionless local
time coordinate $t$ and dimensionless local spatial coordinates
$\{\,x^{i}\,\}$ that comove with ${\bf T}$, the {\em gauge
conditions\/} related to a $1+3$ orthonormal frame representation
are made explicit by \ct{fri96}
\be
e_{0}{}^{\mu} = \frac{1}{T^{0}}\,(M_{0}^{-1}\,\d^{\mu}{}_{0} -
T^{\alpha}\,e_{\alpha}{}^{\mu}) \ , \hsp5
\Gam^{0}{}_{\alpha 0} = \frac{1}{T^{0}}\,(T^{0}{}_{\alpha}
- \Gam^{0}{}_{\alpha\beta}\,T^{\beta}) \ , \hsp5
\Gam^{\alpha}{}_{\beta 0} = \frac{1}{T^{0}}\,
(T^{\alpha}{}_{\beta} - \Gam^{\alpha}{}_{\beta\gam}\,T^{\gam}) \ ;
\ee
we keep the inverse unit of $\lgth$, $M_{0}^{-1}$, as a coefficient
for reasons of physical dimensions.

The fluid-comoving, {\em Lagrangean\/} perspective adopted in
the discussion of Ref. \ct{hveell99} by identifying the timelike
reference congruence with the fluid 4-velocity field, $\p_{0}
\equiv {\bf u}$, is now obtained by fixing three of the four
dimensionless coordinate gauge source functions according to
$T^{\alpha} = 0$, resulting in an alignment ${\bf T} \parallel
\p_{0}$ ($\equiv {\bf u}$). This leads to
\be
e_{0}{}^{\mu} = M^{-1}\,\d^{\mu}{}_{0} \ , \hsp5 
T^{0}{}_{\alpha} = T^{0}\,\Gam^{0}{}_{\alpha 0}
= T^{0}\,\udot_{\alpha} \ , \hsp5
T^{\alpha}{}_{\beta} = T^{0}\,\Gam^{\alpha}{}_{\beta 0}
= T^{0}\,\eps^{\alpha}{}_{\beta\gam}\,\Om^{\gam} \ ,
\ee
where $M := T^{0}M_{0}$. Consequently, the three frame gauge source
functions $T^{0}{}_{\alpha}$ become proportional to the components
of the fluid acceleration $\udot^{\alpha}$, while the three frame
gauge source functions $T^{\alpha}{}_{\beta}$ become proportional
to the components of the rotation rate $\Om^{\alpha}$ at which the
spatial frame $\{\,\p_{\alpha}\,\}$ fails to be Fermi-propagated
along ${\bf u}$. For cosmological models $\left(\,{\cal M},\,{\bf
g},\,{\bf u}\,\right)$ with perfect fluid matter sources,
Ref. \ct{hveell99} introduced {\em proper time\/} along ${\bf u}$
by setting $T^{0} = 1 \Rightarrow M = M_{0}$, and derived the
evolution equation for $\udot^{\alpha}$ along ${\bf u}$ from the
commutators on the basis of Eqs. (\r{eos}) and (\r{mom})
below. Additionally, Ref. \ct{hveell99} set $\Om^{\alpha} = 0$. The
latter choice, however, is by no means compulsory, and other
choices may prove equally convenient (given that $\Om^{\alpha}$ is
assumed to be ``known'').

In order to obtain from the extended $1+3$ orthonormal frame
relations proper partial differential equations such that the
theory underlying Subsec. \r{subsec:fosh} applied,
Ref. \ct{hveell99} expressed the coordinate components
$e_{0}{}^{\mu} := \p_{0}(x^{\mu})$ and $e_{\alpha}{}^{\mu} :=
\p_{\alpha}(x^{\mu})$ of the $1+3$ ONF basis $\{\,\p_{0},
\,\p_{\alpha}\,\}$ in terms of the comoving local coordinate basis
$\{\,\ptl_{t}, \,\ptl_{i}\,\}$ by
\be
\l{onf13}
\p_{0} := M^{-1}\,\ptl_{t} \ , \hsp5
\p_{\alpha} := e_{\alpha}{}^{i}\,(M_{i}\,\ptl_{t}+\ptl_{i}) \ ;
\ee
$M = M(t,x^{i})$ denotes the {\em threading lapse function\/} and
$M_{i}\,dx^{i} = M_{i}(t,x^{j})\,dx^{i}$ the dimensionless {\em
threading shift 1-form\/}. The inverse of the {\em threading
metric\/} is $h^{ij} := \d^{\alpha\beta}\,e_{\alpha}{}^{i}\,
e_{\beta}{}^{j}$. See, e.g., Ref. \ct{hveugg97} and references
therein.

%------------------------------------------------------------------
\subsection{Matter model}
\l{subsec:matter}
%------------------------------------------------------------------
The matter sources in the present discussion (as well as in
Ref. \ct{hveell99}) are assumed to be modelled phenomenologically
as a {\em perfect fluid\/} such that, with respect to
fluid-comoving observers,
\be
0 = q^{\alpha}({\bf u}) = \pi_{\alpha\beta}({\bf u}) \ ,
\ee
i.e., the energy current density and the anisotropic pressure both
vanish. Additionally, a {\em baryotropic\/} equation of state is
assumed,
\be
\l{eos}
p = p(\mu) \ ,
\ee
relating the isotropic pressure $p({\bf u})$ to the total energy
density $\mu({\bf u})$.

%------------------------------------------------------------------
\subsection{Constraint equations}
\l{subsec:cons}
%------------------------------------------------------------------
The following relations in the set obtained from an extended $1+3$
orthonormal frame representation of the relativistic gravitational
field equations do {\em not\/} contain any frame derivatives with
respect to $\p_{0}$. Hence, it is commonplace to refer to these
relations as ``constraint equations''.\footnote{Even though this
terminology is problematic in the generic case when $\p_{0} \equiv
{\bf u}$ has {\em non-zero\/} vorticity, $\om^{\alpha}({\bf u})
\neq 0$, and local coordinates are introduced according to
Eq. (\r{onf13}) above.} These are \ct{mac73,hveugg97}
\bea
\l{onfdivsig}
0 & = & (C_{1})^{\alpha} \ := \ (\p_{\beta} - 3\,a_{\beta})\,
(\sig^{\alpha\beta}) - \sfrac{2}{3}\,\d^{\alpha\beta}\,
\p_{\beta}(\Th) - n^{\alpha}\!_{\beta}\,\om^{\beta}
+ \epsilon^{\alpha\beta\gamma}\,
[\ (\p_{\beta} + 2\,\udot_{\beta} - a_{\beta})\,
(\om_{\gamma}) - n_{\beta\delta}\,\sig^{\delta}\!_{\gamma}\ ]
\\ \nonumber \\
\l{onfdivom}
0 & = & (C_{2}) \ := \ (\p_{\alpha} - \udot_{\alpha}
- 2\,a_{\alpha})\,(\om^{\alpha})
\\ \nonumber \\
\l{onfhconstr}
0 & = & (C_{3})^{\alpha\beta} \ := \ (\d^{\gam\la\alpha}\,
\p_{\gam} + 2\,\udot^{\la\alpha} + a^{\la\alpha})\,(\om^{\beta\ra}) 
- \sfrac{1}{2}\,n^{\gam}\!_{\gam}\,\sig^{\alpha\beta}
+ 3\,n^{\la\alpha}\!_{\gam}\,\sig^{\beta\ra\gamma} 
+ H^{\alpha\beta} \nonumber \\
& & \hspace{25mm} - \ \eps^{\gam\delta\la\alpha}\,[\ (\p_{\gam}
- a_{\gam})\,(\sig^{\beta\ra}\!_{\delta})
+ n^{\beta\ra}\!_{\gam}\,\om_{\delta}\ ]
\\ \nonumber \\
\l{onfjac}
0 & = & (C_{\rm J})^{\alpha} \ := \ (\p_{\beta} - 2\,a_{\beta})\,
(n^{\alpha\beta}) + \sfrac{2}{3}\,\Th\,\om^{\alpha}
+ 2\,\sig^{\alpha}\!_{\beta}\,\om^{\beta}
+ \eps^{\alpha\beta\gam}\,[\ \p_{\beta}(a_{\gam}) - 2\,\om_{\beta}
\,\Omega_{\gam}\ ]
\\ \nonumber \\
\l{onfgauss}
0 & = & (C_{\rm G})^{\alpha\beta} \ := \ {}^{*}\!S^{\alpha\beta}
+ \sfrac{1}{3}\,\Th\,\sig^{\alpha\beta}
- \sig^{\la\alpha}\!_{\gam}\,\sigma^{\beta\ra\gam}
- \om^{\la\alpha}\,\om^{\beta\ra} + 2\,\om^{\la\alpha}\,
\Omega^{\beta\ra} - E^{\alpha\beta}
\\ \nonumber \\
\l{onffried}
0 & = & (C_{\rm G}) \ := \ {}^{*}\!R + \sfrac{2}{3}\,\Th^{2}
- (\sig_{\alpha\beta}\sig^{\alpha\beta})
+ 2\,(\om_{\alpha}\om^{\alpha})
- 4\,(\om_{\alpha}\,\Omega^{\alpha}) - 2\,\mu - 2\,\Lambda
\\ \nonumber \\
\l{onfdive}
0 & = & (C_{4})^{\alpha} \ := \ (\p_{\beta} - 3\,a_{\beta})
\,(E^{\alpha\beta})
- \sfrac{1}{3}\,\d^{\alpha\beta}\,\p_{\beta}(\mu)
- 3\,\om_{\beta}\,H^{\alpha\beta}
- \eps^{\alpha\beta\gam}\,[\ \sig_{\beta\delta}
\,H^{\delta}\!_{\gam} + n_{\beta\delta}\,E^{\delta}\!_{\gam}\ ]
\\ \nonumber \\
\l{onfdivh}
0 & = & (C_{5})^{\alpha} \ := \ (\p_{\beta} - 3\,a_{\beta})\,
(H^{\alpha\beta}) + (\mu+p)\,\om^{\alpha}
+ 3\,\om_{\beta}\,E^{\alpha\beta}
+ \eps^{\alpha\beta\gam}\,[\ \sig_{\beta\delta}\,
E^{\delta}\!_{\gam} - n_{\beta\delta}\,H^{\delta}\!_{\gam}\ ]
\\ \nonumber \\
\l{mom}
0 & = & (C_{\rm PF})^{\alpha} \ := \ \cs^{2}\,\d^{\alpha\beta}\,
\p_{\beta}(\mu) + (\mu+p)\,\udot^{\alpha} \ ,
\eea
where
\bea
\l{onftf3ric}
{}^{*}\!S_{\alpha\beta} & := & \p_{\la\alpha}(a_{\beta\ra})
+ b_{\la\alpha\beta\ra} - \eps^{\gam\delta}{}_{\la\alpha}\,
(\p_{|\gam|} - 2\,a_{|\gam|})\,(n_{\beta\ra\delta})
\\ \nonumber \\
\l{onf3rscl}
{}^{*}\!R & := &  2\,(2\,\p_{\alpha} - 3\,a_{\alpha})\,
(a^{\alpha}) - \sfrac{1}{2}\,b^{\alpha}\!_{\alpha}
\\ \nonumber \\
b_{\alpha\beta} & := & 2\,n_{\alpha\gam}\,n_{\beta}\!^{\gam}
- n^{\gam}\!_{\gam}\,n_{\alpha\beta} \ ,
\eea
$\cs^{2}(\mu) := dp(\mu)/d\mu$ defines the {\em isentropic speed of
sound\/} with $0 \leq \cs^{2} \leq 1$, and angle brackets denote
the symmetric tracefree part.\footnote{The numbering of the
constraint equations we employ is based on the conventions
established in $1+3$ covariant treatments of relativistic
cosmological models $\left(\,{\cal M},\,{\bf g},\,{\bf u}\,\right)$
(cf. Ref. \ct{ellhve99}).} While Eqs. (\r{onfdivsig}) --
(\r{onffried}) derive from the Ricci identities, the Jacobi
identities and the Einstein field equations and so are of order
$\ptl^{2}g$, Eqs. (\r{onfdive}) -- (\r{mom}) derive from the once-
and twice-contracted second Bianchi identities and so are of order
$\ptl^{3}g$. The divergence constraint equation (\r{onfdivsig}) for
the fluid rate of shear is often referred to as the ``momentum
constraint''. When $0 = \om^{\alpha}({\bf u})$, such that the fluid
4-velocity field ${\bf u}$ constitutes the normals to a family of
spacelike 3-surfaces ${\cal S}$:$\{t=\mbox{const}\}$,
Eqs. (\r{onfgauss}) and (\r{onffried}) correspond to the symmetric
tracefree and trace parts of the once-contracted Gau\ss\ embedding
equation. In this case, one also speaks of $(C_{G})$ as the
generalised Friedmann equation, alias the ``Hamiltonian
constraint'' or the ``energy constraint''.

%------------------------------------------------------------------
\subsection{$(1+1+2)$--decomposition}
\l{subsec:112dec}
%------------------------------------------------------------------
In the present discussion it proves very helpful to consider a
$(1+1+2)$--decomposition of all geometrically defined field
variables and their dynamical relations. In order to do so, we
arbitrarily pick the frame basis field $\p_{1}$ as a second,
spacelike, reference direction, in addition to $\p_{0} \equiv {\bf
u}$ as a timelike one; any other spatial direction, however, would
be equally acceptable. Hence, in a small isotropic neighbourhood
${\cal U}$ in the local rest 3-space of an arbitrary event ${\cal
P}$, we establish the convention of regarding those spatial frame
components of geometrical objects which contain the index ``$1$''
as (semi-){\em longitudinal\/} with respect to $\p_{1}$, while
regarding those which exclude the index ``$1$'' as {\em
transverse\/} with respect to $\p_{1}$. Likewise, in ${\cal U}$,
$\p_{1}$ shall constitute the {\em outward\/} frame derivative
while $\p_{2}$ and $\p_{3}$ shall be {\em tangential\/} frame
derivatives when we turn in Sec. \r{sec:jumps} to the issue of jump
discontinuities across spherical spacelike 2-surfaces ${\cal
J}$:$\{t = \mbox{const},\phi(x^{\mu}) = \mbox{const}\}$. For the
frame components of spatial rank-$2$ symmetric tracefree tensors
$a_{\alpha\beta} = a_{\la\alpha\beta\ra}$ with squared magnitude
$a^{2} := \sfrac{1}{2}(a_{\alpha\beta}a^{\alpha\beta})
\ge 0$, we define a {\em new\/} set of frame variables by
\bea
\l{apm}
\begin{array}{lllll}
a_{+} := \sfrac{1}{2}\,(a_{22}+a_{33}) = - \sfrac{1}{2}\,a_{11} & &
a_{-} := \sfrac{1}{2\sqrt{3}}\,(a_{22}-a_{33}) & & \\ \\
a_{\times} := \sfrac{1}{\sqrt{3}}\,a_{23} & &
a_{2} := \sfrac{1}{\sqrt{3}}\,a_{31} & &
a_{3} := \sfrac{1}{\sqrt{3}}\,a_{12} \ ,
\end{array}
\eea
so that
\be
a^{2} = 3\,(a_{+}^{2} + a_{-}^{2} + a_{\times}^{2}
+ a_{2}^{2} + a_{3}^{2}) \ .
\ee
In particular, in the present discussion we have $a_{\alpha\beta}
\in \{\,\sig_{\alpha\beta}, \,E_{\alpha\beta},
\,H_{\alpha\beta}\,\}$. We remark that these definitions are now
adapted to the conventions of the book edited by Wainwright and
Ellis \ct{waiell97}, implying they differ by a factor of
$\sfrac{1}{3}$ from those used in Ref. \ct{hveell99}. In analogy to
Eq. (\r{apm}), we perform a $(1+1+2)$--decomposition of the spatial
commutation functions $n_{\alpha\beta}$ by defining
\bea
\l{npm}
\begin{array}{lllll}
n := n_{11} + n_{22} + n_{33} & &
\np := -\,n_{11} + \sfrac{1}{2}\,(n_{22}+n_{33}) & &
\nm := \sfrac{1}{2\sqrt{3}}\,(n_{22}-n_{33}) \\ \\
\nc := \sfrac{1}{\sqrt{3}}\,n_{23} & &
\ntw := \sfrac{1}{\sqrt{3}}\,n_{31} & &
\nth := \sfrac{1}{\sqrt{3}}\,n_{12} \ .
\end{array}
\eea
The squared magnitude is then given by
\be
\sfrac{1}{2}(n_{\alpha\beta}n^{\alpha\beta})
= \sfrac{1}{6}\,(n^{2} + 2\np^{2})
+ 3\,(\nm^{2} + \nc^{2} + \ntw^{2} + \nth^{2}) \ .
\ee
Note that only $(n-2\np)$, $\nm$ and $\nc$ transform as {\em
tensor components\/} under rotations of the spatial frame
$\{\,\p_{\alpha}\,\}$ about the reference $\p_{1}$-axis.

By employing these conventions and definitions, we have listed in
Appendix \r{app2} certain linear combinations of the components of
the constraint equations (\r{onfdivsig}) -- (\r{mom}) that will be
needed in the following sections.

\begin{center}
{\bf *************************************************************}

INSERT Table \r{tab:112conv} on ``Conventions for
$(1+1+2)$--decomposition'' HERE.

{\bf *************************************************************}
\end{center}
%

%%%%%%%%%%%%%%%%%%%%%%%%%%%%%%%%%%%%%%%%%%%%%%%%%%%%%%%%%%%%%%%%%%%
\section{Physical effect of constraint equations on outward first
derivatives}
\l{sec:jumps}
%%%%%%%%%%%%%%%%%%%%%%%%%%%%%%%%%%%%%%%%%%%%%%%%%%%%%%%%%%%%%%%%%%%
Generic cosmological models $\left(\,{\cal M},\,{\bf g},\,{\bf
u}\,\right)$ with a perfect fluid matter source have fluid
4-velocity fields ${\bf u}$ with {\em non-zero\/} vorticity,
$\om^{\alpha}({\bf u}) \neq 0$. This property makes it impossible
to determine a fluid-comoving spacelike 3-surface ${\cal
S}$:$\{t=\mbox{const}\}$ {\em everywhere\/} orthogonal to ${\bf u}$
on which initial data satisfying the constraint equations of
Subsec. \r{subsec:cons} could be specified. In this case, the
discussion of a well-posed Cauchy initial value problem requires
that the setting of the data as well as the solution of the
constraint equations be instead performed on a {\em non-comoving\/}
spacelike 3-surface ${\cal S}$:$\{t=\mbox{const}\}$, before the
data is evolved along ${\bf u}$.\footnote{It is currently unknown
whether the $\ptl^{3}g$-order FOSH evolution systems with perfect
fluid matter sources in Refs. \ct{fri98} and \ct{hveell99} can be
generalised to a non-comoving perspective.} All these complications
disappear when $0 = \om^{\alpha}({\bf u})$, and well-defined
spacelike 3-surface ${\cal S}$:$\{t=\mbox{const}\}$ everywhere
orthogonal to ${\bf u}$ {\em do\/} exist.

For our purposes, however, it is sufficient to investigate the
physical effect of the constraint equations on the selection of
appropriate initial data sets from a purely {\em local\/}
viewpoint, i.e., only in a small isotropic neighbourhood ${\cal U}$
in the local rest 3-space of an arbitrary event ${\cal P}$. Due to
the local Minkowskian structure of all relativistic spacetime
manifolds $\left(\,{\cal M},\,{\bf g}\,\right)$, one conventionally
determines (and analyses their physical properties) the set of
characteristic cones ${\cal C}$:$\{\phi(x^{\mu}) =\mbox{const}\}$
for a given FOSH evolution system only within the small isotropic
spacetime neighbourhood $\{-\,\varepsilon \leq t \leq \varepsilon\}
\times {\cal U}$ of ${\cal P}$. In line with this, we will consider in
the following spherical spacelike 2-surfaces ${\cal J}$:$\{t =
\mbox{const}, \phi(x^{\mu}) = \mbox{const}\}$ in ${\cal U}$ across
which we assume to exist (i) {\em jump discontinuities\/} in the
{\em outward first frame derivatives\/} of certain geometrically
defined field variables $u^{A}$, and (ii) {\em continuity\/} of the
tangential first frame derivatives of the $u^{A}$ and the $u^{A}$
themselves. As the constraint equations have to be satisfied {\em
everywhere\/}, it is clear that across ${\cal J}$ we have $0 =
[(C)^{\alpha\dots}]_{t,\phi = {\rm const}}$ for the value of any
component in the set of Eqs. (\r{onfdivsig}) -- (\r{mom}).

Motivated by the prospect of grounding the discussion on wave-like
phenomena described by the relativistic gravitational field
equations on the deviation equation for a set of test particles,
the dynamical considerations on the $\ptl^{3}g$-order FOSH
evolution system in Ref. \ct{hveell99} focused on the set of Weyl
curvature characteristic eigenfields $\{\,E_{+}, \,H_{+},
\,(E_{3}\mp H_{2}), \,(E_{2}\pm H_{3}), \,(E_{-}\mp H_{\times}),
\,(E_{\times}\pm H_{-})\,\}$. There it was correctly argued that
the deviation equation monitors the physical effects on the state
of motion of a set of test particles of both {\em gradual\/} as
well as {\em sudden\/} changes in the values of these fields (see,
e.g., Refs. \ct{pir56} and \ct{sze65}). What was overlooked in this
work, however, is the fact that, on the basis of the theory
underlying Subsec. \r{subsec:fosh}, the $\ptl^{3}g$-order FOSH
evolution system presented in Ref. \ct{hveell99} can at best
describe the physical transport along bicharacteristic rays of jump
discontinuities in the (outward) first {\em derivatives\/} of these
fields rather than these fields themselves, given the constraint
equations do not impose any additional restrictions. It is our aim
to supplement the discussion of Ref. \ct{hveell99} by such a
consideration in the present section.

%------------------------------------------------------------------
\subsection{Jump discontinuities at derivative level $\ptl^{3}g$}
%------------------------------------------------------------------
Considering in ${\cal U}$ a spherical spacelike 2-surface ${\cal
J}$:$\{t = \mbox{const}, \phi(x^{\mu}) = \mbox{const}\}$, and
assuming that across ${\cal J}$ all geometrically defined field
variables $u^{A}$ as well as their tangential first frame
derivatives are continuous, we find that the Weyl curvature
divergence equations (\r{c41}) -- (\r{c43c52}) amongst the
constraint equations lead to the following set of jump conditions:
\bea
\left[\p_{1}(E_{+})\right]_{t,\phi = {\rm const}}
& = & -\,\sfrac{1}{6}
\left[\p_{1}(\mu)\right]_{t,\phi = {\rm const}} \\
\left[\p_{1}(H_{+})\right]_{t,\phi = {\rm const}}
& = & 0 \\
\l{noj1}
\left[\p_{1}(E_{3}\mp H_{2})\right]_{t,\phi = {\rm const}}
& = & 0 \\
\l{noj2}
\left[\p_{1}(E_{2}\pm H_{3})\right]_{t,\phi = {\rm const}}
& = & 0 \\
\left[\p_{1}(E_{-}\mp H_{\times})\right]_{t,\phi = {\rm const}}
& = & \mbox{unconstrained} \\
\left[\p_{1}(E_{\times}\pm H_{-})\right]_{t,\phi = {\rm const}}
& = & \mbox{unconstrained} \ .
\eea
The implications are three-fold: \enl

(i) Jump discontinuities in the outward first frame derivative of
the Coulomb-like Weyl curvature characteristic eigenfield $E_{+}$
originate from jump discontinuities in the outward first frame
derivative of the matter total energy density $\mu$ and are
physically allowed, if the momentum conservation equation
(\r{bimom1}) is satisfied on both sides of ${\cal J}$ with
$0 = [(C_{\rm PF})_{1}]_{t,\phi = {\rm const}}$. In more detail:
assuming that all of $p$, $\cs$ and $\udot_{1}$ are continuous
across ${\cal J}$, Eq. (\r{bimom1}) yields
\be
\l{cpfjump}
0 = [(C_{\rm PF})_{1}]_{t,\phi = {\rm const}}
= \cs^{2}\,\left[\p_{1}(\mu)\right]_{t,\phi = {\rm const}}
+ \left[\mu\right]_{t,\phi = {\rm const}}\,\udot_{1} \ .
\ee
Hence, if $p = 0 \Rightarrow \cs = 0$ and $\udot_{1} = 0$ on both
sides of ${\cal J}$, the values of $[\p_{1}(\mu)]_{t,\phi = {\rm
const}}$ and $[\mu]_{t,\phi = {\rm const}}$ remain
unconstrained. Phenomena of the presently described kind occur, for
example, across the surfaces of static, spherically symmetrical
perfect fluid stars with equation of state (\r{eos}) (see, e.g.,
Ref. \ct{mtw73}). In a $\ptl^{3}g$-order formulation, it follows
that real-valued initial data for $\mu$ (and so for $E_{+}$) is
required to be of differentiability class $C^{2}({\cal U})$ with
respect to the zeroth-order derivative level of ${\bf g}$. The
equations of Ref. \ct{hveell99} show that $[\p_{1}(E_{+})]_{t,\phi
= {\rm const}} \neq 0$ propagates with characteristic velocity $v =
0$ relative to ${\bf u}$. Note that in the vacuum subcase $\mu = 0
\Rightarrow [\p_{1}(E_{+})]_{t,\phi = {\rm const}} = 0$. Jump
discontinuities in $\p_{1}(H_{+})$ are not physically allowed, and
so, in a $\ptl^{3}g$-order formulation, real-valued initial data
for $H_{+}$ is required to be of differentiability class
$C^{3}({\cal U})$ with respect to the zeroth-order derivative level
of ${\bf g}$. \enl

(ii) Jump discontinuities in the outward first frame derivatives of
the semi-longitudinal Weyl curvature characteristic eigenfields
$(E_{3}\mp H_{2})$ and $(E_{2}\pm H_{3})$, that by restricting to
the net $\ptl^{3}g$-order FOSH evolution system in
Ref. \ct{hveell99} ({\em without\/} accounting for the constraint
equations) are theoretically associated with characteristic
velocities $v = \pm\,\sfrac{1}{2}$ relative to ${\bf u}$, are {\em
not physically allowed\/}. Hence, in a $\ptl^{3}g$-order
formulation, real-valued initial data for $(E_{3}\mp H_{2})$ and
$(E_{2}\pm H_{3})$ needs to be of differentiability class
$C^{3}({\cal U})$ (rather than $C^{2}({\cal U})$) with respect to
the zeroth-order derivative level of ${\bf g}$. In short, {\em not
all\/} initial data can be given freely.

It can be easily inferred from the propagation equations along
${\bf u}$ for the Weyl curvature divergence equations (\r{onfdive})
and (\r{onfdivh}), first published for an irrotational
pressure-free fluid matter source in Ref. \ct{maa97}, and presented
for a general perfect fluid matter source in Refs. \ct{hve98} and
\ct{friren00}, that the characteristic velocities relative to ${\bf
u}$ for the components $(C_{4})_{2}\mp (C_{5})_{3}$ and
$(C_{4})_{3}\pm (C_{5})_{2}$ are $v = \pm\,\sfrac{1}{2}$
too. Hence, comparing this result with Eqs. (\r{c42c53}) and
(\r{c43c52}) in Appendix \r{app2} and Eqs. (\r{noj1}) and
(\r{noj2}) above, it becomes clear that the Weyl curvature
divergence equations propagate relative to ${\bf u}$ at {\em
precisely\/} the speed that is required to ensure that jump
discontinuities in $\p_{1}(E_{3}\mp H_{2})$ and $\p_{1}(E_{2}\pm
H_{3})$ will {\em always remain physically disallowed\/} at any
instant throughout the dynamical evolution of a cosmological model
$\left(\,{\cal M},\,{\bf g},\,{\bf u}\,\right)$. It should be
emphasised at this stage that this property is completely
independent of the presence of matter. That is, of course the jump
conditions (\r{noj1}) and (\r{noj2}) apply equally to vacuum
spacetime configurations. \enl

(iii) Jump discontinuities in the outward first frame derivatives
of the transverse Weyl curvature characteristic eigenfields
$(E_{-}\mp H_{\times})$ and $(E_{\times}\pm H_{-})$ are physically
allowed. Clearly, this situation reflects the freedom of specifying
four {\em arbitrary\/} (non-analytic) real-valued functions
$I_{\ptl^{3}g} := \{\,a_{1}(x^{i}), \,a_{2}(x^{i}), \,a_{3}(x^{i}),
\,a_{4}(x^{i})\,\}$ of differentiability class $C^{2}({\cal U})$
with respect to the zeroth-order derivative level of ${\bf g}$ as
the initial data for the dynamical degrees of freedom associated
with the gravitational field itself.

%------------------------------------------------------------------
\subsection{Jump discontinuities at derivative level $\ptl^{2}g$}
%------------------------------------------------------------------
To be able to argue in terms of physical effects described by the
deviation equation for a set of test particles, we have to turn our
attention directly to the set of Weyl curvature characteristic
eigenfields and possible discontinuous changes in their
values. Such changes are driven by the dynamics of the underlying
connection fields and their derivatives. Hence, to facilitate the
interpretation of generic gravitational dynamics, it would be
desirable to have available a $\ptl^{2}g$-order FOSH evolution
system derived from a $1+3$ orthonormal frame {\em connection\/}
formulation of gravitational fields (see Refs. \ct{mac73} and
\ct{hveugg97} for reviews of the latter). Unfortunately, to date,
such a formulation has not been accomplished for the generic
case. The exception is the $\ptl^{2}g$-order FOSH evolution system
for perfect fluid cosmological models $\left(\,{\cal M},\,{\bf
g},\,{\bf u}\,\right)$ with an Abelian $G_{2}$ isometry group we
will present in Sec. \r{sec:g2fosh} below. In the absence of such a
generally applicable dynamical formulation, we return to our local
viewpoint and investigate how, in a small isotropic neighbourhood
${\cal U}$ in the local rest 3-space of an arbitrary event ${\cal
P}$, the constraint equations at derivative level $\ptl^{2}g$
restrict the occurrence of jump discontinuities in the values of
the Weyl curvature characteristic eigenfields themselves. To this
end, we now focus on Eqs. (\r{onfdivsig}) -- (\r{onffried}), and
certain linear combinations of the components thereof provided by
Eqs. (\r{c11}) -- (\r{cfried}) and (\r{elep}) -- (\r{tweyl2}).
Again, we consider in ${\cal U}$ a spherical spacelike 2-surface
${\cal J}$:$\{t = \mbox{const}, \phi(x^{\mu}) = \mbox{const}\}$,
and assume that across ${\cal J}$ all geometrically defined field
variables $u^{A}$ as well as their tangential first frame
derivatives are continuous. This leads to: \enl

\noindent
(i) {\em $v = 0$ longitudinal Weyl curvature characteristic
eigenfields\/}: \nopagebreak
\bea
E_{+} & = & -\,\sfrac{1}{3}\,\p_{1}(a_{1})
+ \mbox{tangential frame derivatives/algebraic terms} \\
H_{+} & = & \sfrac{1}{3}\,\p_{1}(\om_{1})
+ \mbox{tangential frame derivatives/algebraic terms} \ ,
\eea
from Eqs. (\r{elep}) and (\r{magp}). Across ${\cal J}$, the
generalised Gau\ss--Friedmann equation (\r{cfried}) and the fluid
vorticity divergence equation (\r{cdivom}), respectively, then
impose the restrictions
\bea
\left[E_{+}\right]_{t,\phi = {\rm const}}
& = & -\,\sfrac{1}{3}\left[\p_{1}(a_{1})
\right]_{t,\phi = {\rm const}}
= -\,\sfrac{1}{6}\left[\mu\right]_{t,\phi = {\rm const}} \\
\left[H_{+}\right]_{t,\phi = {\rm const}}
& = & \sfrac{1}{3}\left[\p_{1}(\om_{1})
\right]_{t,\phi = {\rm const}} = 0 \ .
\eea
That is, jump discontinuities in the values of the Coulomb-like
Weyl curvature characteristic eigenfield $E_{+}$ originate from
jump discontinuities in the values of the matter total energy
density $\mu$ and are physically allowed, if the momentum
conservation equation (\r{bimom1}) is satisfied on both sides of
${\cal J}$ with $0 = [(C_{\rm PF})_{1}]_{t,\phi = {\rm const}}$
[cf. Eq. (\r{cpfjump})]. In a $\ptl^{2}g$-order formulation,
real-valued initial data for $\mu$ (and so for $E_{+}$) is thus
required to be of differentiability class $C^{1}({\cal U})$ with
respect to the zeroth-order derivative level of ${\bf g}$. Note
that in the vacuum subcase $\mu = 0 \Rightarrow [E_{+}]_{t,\phi =
{\rm const}} = 0$. Jump discontinuities in the values of $H_{+}$
are not physically allowed, and so, in a $\ptl^{2}g$-order
formulation, real-valued initial data for $H_{+}$ is required to be
of differentiability class $C^{2}({\cal U})$ with respect to the
zeroth-order derivative level of ${\bf g}$. \enl

\noindent
(ii) {\em $v = \pm\,\sfrac{1}{2}$ semi-longitudinal Weyl curvature
characteristic eigenfields\/}: \nopagebreak
\bea
(E_{3}\mp H_{2})
& = & \mp\,\sfrac{1}{2}\,\p_{1}(\sigth\mp\ntw-\sfrac{1}{\sqrt{3}}\, 
\om_{3}\mp\sfrac{1}{\sqrt{3}}\,a_{2})
+ \mbox{tangential frame derivatives/algebraic terms} \\
(E_{2}\pm H_{3})
& = & \mp\,\sfrac{1}{2}\,\p_{1}(\sigtw\pm\nth+\sfrac{1}{\sqrt{3}}\, 
\om_{2}\mp\sfrac{1}{\sqrt{3}}\,a_{3})
+ \mbox{tangential frame derivatives/algebraic terms} \ ,
\eea
from Eqs. (\r{lweyl1}) and (\r{lweyl2}). In this case, we find that
across ${\cal J}$ the fluid shear divergence/Jacobi constraint
equations (\r{c12jac3}) and (\r{c13jac2}) impose the restrictions
\bea
\l{nolwj1}
\left[(E_{3}\mp H_{2})\right]_{t,\phi = {\rm const}}
& = & \mp\,\sfrac{1}{2}\,[\p_{1}(\sigth\mp\ntw
-\sfrac{1}{\sqrt{3}}\,\om_{3}
\mp\sfrac{1}{\sqrt{3}}\,a_{2})]_{t,\phi = {\rm const}} = 0 \\
\l{nolwj2}
\left[(E_{2}\pm H_{3})\right]_{t,\phi = {\rm const}}
& = & \mp\,\sfrac{1}{2}\,[\p_{1}(\sigtw\pm\nth
+\sfrac{1}{\sqrt{3}}\,\om_{2}
\mp\sfrac{1}{\sqrt{3}}\,a_{3})]_{t,\phi = {\rm const}} = 0 \ .
\eea
That is, jump discontinuities in the values of the
semi-longitudinal Weyl curvature characteristic eigenfields
$(E_{3}\mp H_{2})$ and $(E_{2}\pm H_{3})$ are {\em not physically
allowed\/}. In a $\ptl^{2}g$-order formulation real-valued initial
data for $(E_{3}\mp H_{2})$ and $(E_{2}\pm H_{3})$ is thus required
to be of differentiability class $C^{2}({\cal U})$ (rather than
$C^{1}({\cal U})$) with respect to the zeroth-order derivative
level of ${\bf g}$. \enl

\noindent
(iii) {\em $v = \pm\,1$ transverse Weyl curvature characteristic
eigenfields\/}: \nopagebreak
\bea
(E_{-}\mp H_{\times})
& = & \mp\,\p_{1}(\sigm\mp\nc)
+ \mbox{tangential frame derivatives/algebraic terms} \\
(E_{\times}\pm H_{-})
& = & \mp\,\p_{1}(\sigc\pm\nm)
+ \mbox{tangential frame derivatives/algebraic terms} \ ,
\eea
from Eqs. (\r{tweyl1}) and (\r{tweyl2}). In this case, we find that
across ${\cal J}$ the constraint equations impose {\em no\/}
restrictions so that
\bea
\left[(E_{-}\mp H_{\times})\right]_{t,\phi = {\rm const}}
& = & \mp\left[\p_{1}(\sigm\mp\nc)\right]_{t,\phi = {\rm const}}
= \mbox{unconstrained} \\
\left[(E_{\times}\pm H_{-})\right]_{t,\phi = {\rm const}}
& = & \mp\left[\p_{1}(\sigc\pm\nm)\right]_{t,\phi = {\rm const}}
= \mbox{unconstrained} \ .
\eea
That is, jump discontinuities in the values of the transverse Weyl
curvature characteristic eigenfields $(E_{-}\mp H_{\times})$ and
$(E_{\times}\pm H_{-})$ are physically allowed. Hence, they can
transport arbitrary non-zero jump discontinuities
$[\p_{1}(\sigm\mp\nc)]_{t,\phi = {\rm const}}$ and
$[\p_{1}(\sigc\pm\nm)]_{t,\phi = {\rm const}}$ of finite
magnitude. Again, this reflects the freedom of specifying four {\em
arbitrary\/} (non-analytic) real-valued functions $I_{\ptl^{2}g} :=
\{\,A_{1}(x^{i}), \,A_{2}(x^{i}), \,A_{3}(x^{i}),
\,A_{4}(x^{i})\,\}$ of differentiability class $C^{1}({\cal U})$
with respect to the zeroth-order derivative level of ${\bf g}$ as
the initial data for the dynamical degrees of freedom associated
with the gravitational field itself. \enl
%{\bf [\ hve: Issue of possibility of propagating arbitrary
%information at $|\,v\,| = 1$.\ ]}

For completeness, we now also briefly discuss the effect of the
constraint equations (\r{onfdivsig}) -- (\r{onfjac}), respectively,
Eqs. (\r{c11}) -- (\r{cdivom}), on the outward first frame
derivatives of the characteristic eigenfields associated with the
fluid kinematical branch of the $\ptl^{3}g$-order FOSH evolution
system in Ref. \ct{hveell99} [cf. Eqs. (3.27) and (3.28)]. The
derivative level is hence one below the Weyl curvature case. For
the $v = 0$ fluid kinematical characteristic eigenfields we find
the jump conditions: \enl

\noindent
(iv) {\em $v = 0$ fluid kinematical characteristic eigenfields\/}:
\nopagebreak
\bea
& & [\p_{1}(\sfrac{1}{3}\,\Th+\sigp)]_{t,\phi = {\rm const}}
= 0 \\
& & [\p_{1}(\om_{1})]_{t,\phi = {\rm const}}
= 0 \\
\l{rotmod1}
& & [\p_{1}(\sigth+\sfrac{1}{\sqrt{3}}\,\om_{3})
]_{t,\phi = {\rm const}}
= \mbox{unconstrained} \\
\l{rotmod2}
& & [\p_{1}(\sigtw-\sfrac{1}{\sqrt{3}}\,\om_{2})
]_{t,\phi = {\rm const}}
= \mbox{unconstrained} \ .
\eea
The last two conditions imply the existence of two generically
non-zero {\em fluid rotational modes\/} that were identified before
by Ehlers et al in an analysis of linearised perturbations of
arbitrary background dust spacetimes \ct{ehletal87}. Corresponding
real-valued initial data for these modes is required to be of
differentiability class $C^{1}({\cal U})$ with respect to the
zeroth-order derivative level of ${\bf g}$. Finally, for the
different parts of the $v = \pm\,\cs$ fluid kinematical
characteristic eigenfields we find the jump conditions: \enl

\noindent
(v) {\em $v = \pm\,\cs$ fluid kinematical characteristic
eigenfields\/}: \nopagebreak
\bea
& & [\p_{1}(\sfrac{1}{3}\,\Th-2\,\sigp)
]_{t,\phi = {\rm const}}
= \mbox{unconstrained} \\
\l{smod2}
& & [\p_{1}(\sigth-\sfrac{1}{\sqrt{3}}\,\om_{3})
]_{t,\phi = {\rm const}}
= \pm\,[\p_{1}(\sfrac{1}{\sqrt{3}}\,a_{2}+\ntw)
]_{t,\phi = {\rm const}} = \mbox{unconstrained} \\
\l{smod3}
& & [\p_{1}(\sigtw+\sfrac{1}{\sqrt{3}}\,\om_{2})
]_{t,\phi = {\rm const}}
= \pm\,[\p_{1}(\sfrac{1}{\sqrt{3}}\,a_{3}-\nth)
]_{t,\phi = {\rm const}} = \mbox{unconstrained} \\
& & [\p_{1}(\udot_{1})]_{t,\phi = {\rm const}}
= \mbox{unconstrained} \\
& & [\p_{1}(\udot_{2})]_{t,\phi = {\rm const}}
= \mbox{unconstrained} \\
& & [\p_{1}(\udot_{3})]_{t,\phi = {\rm const}}
= \mbox{unconstrained} \ .
\eea
Again, corresponding real-valued initial data for these parts is
required to be of differentiability class $C^{1}({\cal U})$ with
respect to the zeroth-order derivative level of ${\bf g}$. Note
especially that the jump conditions (\r{smod2}) and (\r{smod3}) are
precisely of such a nature that {\em no\/} violations of the jump
conditions (\r{nolwj1}) and (\r{nolwj2}) above for the
semi-longitudinal Weyl curvature characteristic eigenfields may
occur.

%%%%%%%%%%%%%%%%%%%%%%%%%%%%%%%%%%%%%%%%%%%%%%%%%%%%%%%%%%%%%%%%%%%
\section{Worked example: Cosmological models with Abelian $G_{2}$
isometry group}
\l{sec:g2fosh}
%%%%%%%%%%%%%%%%%%%%%%%%%%%%%%%%%%%%%%%%%%%%%%%%%%%%%%%%%%%%%%%%%%%
In this section we turn to discuss in some detail a {\em new\/},
fully gauge-fixed, $\ptl^{2}g$-order autonomous FOSH evolution
system for spatially inhomogeneous perfect fluid cosmological
models $\left(\,{\cal M},\,{\bf g},\,{\bf u}\,\right)$ which are
invariant under the transformations of an {\em Abelian $G_{2}$
isometry group\/} that is simply transitive on spacelike
2-surfaces. Thus, all geometrically defined field variables $u^{A}$
vary in {\em one\/} spatial direction only. A systematic approach
to this class of cosmological models was brought forward some time
ago by Wainwright in Refs. \ct{wai79} and \ct{wai81}, wherein the
generic case was given the classification label ``A(i)''. A number
of exact, real analytic, solutions to the EFE for Abelian $G_{2}$
perfect fluid cosmological models are known, such as those listed
in Refs. \ct{wai81} and \ct{waiell97} or the singularity-free
solution obtained by Senovilla \ct{sen90}, but most of them belong
to dynamically restricted or higher-symmetry subcases.

%------------------------------------------------------------------
\subsection{Well-posed Cauchy initial value problem}
%------------------------------------------------------------------
Choosing an {\em orbit-aligned group-invariant\/} $1+3$ ONF basis
$\{\,\p_{0}, \,\p_{\alpha}\,\}$ such that commutation relations
\be
0 = [\,\mbox{\boldmath $\xi$}, \p_{0}\,]
= [\,\mbox{\boldmath $\xi$}, \p_{\alpha}\,]
= [\,\mbox{\boldmath $\eta$}, \p_{0}\,]
= [\,\mbox{\boldmath $\eta$}, \p_{\alpha}\,]
\ee
hold between the two commuting spacelike {\em Killing vector
fields\/} $\mbox{\boldmath $\xi$}$ and $\mbox{\boldmath $\eta$}$
and $\{\,\p_{0}, \,\p_{\alpha}\,\}$, and assuming that $\p_{0}
\equiv {\bf u}$ is {\em orthogonal\/} to the isometry group orbits,
it follows that for all solutions in the Abelian $G_{2}$ class we
have \ct{wai79,wai81}
\be
\l{zom}
0 = \p_{2}(u^{A}) = \p_{3}(u^{A}) \ , \hsp5
0 = \udot_{2} = \udot_{3} = \om^{\alpha}({\bf u})
= a_{2} = a_{3} = (n-2\np) = \ntw = \nth \ .
\ee
That is, $\p_{2}$ and $\p_{3}$ are tangent to the isometry group
orbits. Besides $\p_{0} \equiv {\bf u}$ also the frame basis field
$\p_{1}$ is {\em hypersurface orthogonal\/} \ct{wai79}. The {\em
canonical choice\/} Wainwright proposes for Abelian $G_{2}$ perfect
fluid cosmological models consists of introducing fluid-comoving
local coordinates $\{\,t, \,x, \,y, \,z\,\}$ adapted to the
integral curves of ${\bf u}$, $\mbox{\boldmath $\xi$}$ and
$\mbox{\boldmath $\eta$}$ such that \ct{wai81}
\be
\mbox{\boldmath $\xi$} = \ptl_{y} \ , \hsp5
\mbox{\boldmath $\eta$} = \ptl_{z} \ ;
\ee
the isometry group orbits, which have vanishing Gau\ss ian
2-curvature, are thus given by spacelike 2-surfaces
$\{t=\mbox{const}, x=\mbox{const}\}$. Additionally, the coordinate
components of the $1+3$ ONF basis $\{\,\p_{0}, \,\p_{\alpha}\,\}$
as introduced by Eq. (\r{onf13}) are specialised to \ct{wai81}
\bea
\l{g2frame}
\begin{array}{cclcccl}
\p_{0} & = & M^{-1}\,\ptl_{t} \ , & \hsp5 &
\p_{2} & = & e_{2}{}^{2}\,\ptl_{y} \ ,\\
\p_{1} & = & e_{1}{}^{1}\,\ptl_{x} + e_{1}{}^{2}\,\ptl_{y}
+ e_{1}{}^{3}\,\ptl_{z}  \ , & \hsp5 & 
\p_{3} & = & e_{3}{}^{2}\,\ptl_{y} + e_{3}{}^{3}\,\ptl_{z} \ ;
\end{array}
\eea
in view of $0 = \om^{\alpha}({\bf u})$ the choice $0 = M_{i}$ is
made.\footnote{For $0 = \om^{\alpha}({\bf u})$, the choice $0 =
M_{i}$ is the simplest one possible, but it is by no means
compulsory.} Note that Wainwright's canonical choice establishes
the property $\p_{2} \parallel \mbox{\boldmath $\xi$}$. All
geometrically defined field variables $u^{A}$ will now only be
functions of the local coordinates $t$ and $x$. From the
commutators (see, e.g., Refs. \ct{mac73} and \ct{hveugg97}), the
canonical choice of $1+3$ ONF basis and local coordinates has the
direct consequences
\be
\l{ccond1}
0 = (\sqrt{3}\,\sigc+\Om_{1})
= (\sqrt{3}\,\sigtw+\Om_{2})
= (\sqrt{3}\,\sigth-\Om_{3}) \ ,
\ee
implying the three spatial coordinate conditions $0 = e_{2}{}^{1} =
e_{2}{}^{3} = e_{3}{}^{1}$ in Eq. (\r{g2frame}) will be
automatically preserved along ${\bf u}$. Thus, the spatial frame
$\{\,\p_{\alpha}\,\}$ will presently {\em not\/} be
Fermi-transported along ${\bf u}$, but $\Om^{\alpha}$ (and so the
three frame gauge source functions $T^{\alpha}{}_{\beta}$) will be
adapted to the fluid rate of shear instead. Likewise, the canonical
choice leads to
\be
\l{ccond2}
0 = (n_{+}-\sqrt{3}\,n_{-}) \ .
\ee
We now define two new frame variables for two components of the
fluid rate of expansion tensor by
\be
\alpha := (\sfrac{1}{3}\,\Th-2\,\sig_{+}) \ ,  \hsp5
\beta := (\sfrac{1}{3}\,\Th+\sig_{+}) \ ,
\ee
shadowing the variable names used earlier in a related context in
Ref. \ct{ell67} (on this choice of variables, see also the remarks
made on the so-called ``Taub gauge'' for fluid spacetime geometries
in Ref. \ct{uggetal95}).

Substituting from Eq. (\r{ccond1}) into the evolution equation for
the semi-longitudinal fluid shear component $\sig_{3}$ (see, e.g.,
Ref. \ct{hveugg97}), one finds that the latter is {\em
involutive\/}. Therefore, on performing a spatial rotation of
$\{\,\p_{\alpha}\,\}$ about $\p_{1}$ at every point of a given
2-surfaces $\{t=\mbox{const}, x=\mbox{const}\}$, one can set
\be
0 = \sig_{3}
\ee
to hold at every event of $\left({\cal M},\,{\bf g},\,{\bf
u}\right)$.\footnote{Alternatively one could choose $\p_{3}
\parallel \mbox{\boldmath $\eta$}$, and then set $0 =
\sig_{2}$ instead.} Determining the evolution along ${\bf u}$ of
the frame gauge source function $(T^{0})^{-1}\,T^{0}{}_{1} =
\udot_{1}$ as outlined in Subsec. \r{subsec:gauge}, at this point,
from a dynamical viewpoint, the coordinate and spatial frame
freedom has been completely fixed (modulo coordinate
reparameterisation freedom given by $t^{\prime} = t^{\prime}(t)$,
$x^{\prime} = x^{\prime}(x)$, $y^{\prime} = y + f(x)$ and
$z^{\prime} = z + g(x)$
\ct{wai81}).

It is now fairly straightforward to derive from the equations of a
$1+3$ orthonormal frame {\em connection\/} formulation of
gravitational fields as given in Refs. \ct{mac73} and \ct{hveugg97}
an evolution system of autonomous partial differential equations in
FOSH format for the following set of eleven geometrically defined
field variables $u^{A} = u^{A}(t,x)$:
\be
\l{depvar}
u^{A} = \left(\,e_{1}{}^{1}, \,\beta,\, \sig_{2}, \,a_{1}, \,\mu,
\,\alpha, \,\udot_{1}, \,\sig_{-}, \,n_{\times},
\,\sig_{\times}, \,n_{-}\,\right)^{T} \ .
\ee
Note that the evolution of the frame coordinate components other
than $e_{1}{}^{1}$ is decoupled from this set; $e_{1}{}^{1}$ itself
forms a background field for pairwise dynamical interactions
between $\{\,\alpha, \,\udot_{1}\,\}$, $\{\,\sigm, \,\nc\,\}$ and
$\{\,\sigc, \,\nm\,\}$ in the sense of Geroch \ct{ger96}. \enl

%\pagebreak
\noindent
{\em 11-dimensional autonomous first-order symmetric hyperbolic
evolution system\/}: \nopagebreak
\bea
\l{e11dot}
M^{-1}\,\ptl_{t}e_{1}{}^{1}
& = & -\,\alpha\,e_{1}{}^{1} \\
\l{betadot}
M^{-1}\,\ptl_{t}\beta
& = & -\,\sfrac{3}{2}\,\beta^{2} - \sfrac{3}{2}\,(\sig_{-}^{2}
+\sig_{\times}^{2}+n_{-}^{2}+n_{\times}^{2})
+ \sfrac{3}{2}\,\sig_{2}^{2}
- \sfrac{1}{2}\,(2\,\udot_{1}-a_{1})\,a_{1}
- \sfrac{1}{2}\,(p-\Lambda) \\
\l{sig2dot}
M^{-1}\,\ptl_{t}\sig_{2}
& = & -\,(3\,\beta-\sqrt{3}\,\sig_{-})\,\sig_{2} \\
\l{a1dot}
M^{-1}\,\ptl_{t}a_{1}
& = & -\,\beta\,(\udot_{1}+a_{1}) + 3\,(n_{-}\,\sig_{\times}
-n_{\times}\,\sig_{-}) \\
\l{mudot}
M^{-1}\,\ptl_{t}\mu
& = & -\,(\alpha+2\,\beta)\,(\mu+p) \\
\l{alphadot}
\cs^{2}\,M^{-1}\,\ptl_{t}\alpha - \cs^{2}\,e_{1}{}^{1}\,
\ptl_{x}\udot_{1}
& = & \cs^{2}\,[\ -\,\alpha^{2} + \beta^{2} - 3\,(\sig_{-}^{2}
+\sig_{\times}^{2}-n_{-}^{2}-n_{\times}^{2}) - 9\,\sig_{2}^{2}
+ \udot_{1}^{2} - a_{1}^{2} - \sfrac{1}{2}\,(\mu+p)\ ] \\
\l{udot1dot}
M^{-1}\,\ptl_{t}\udot_{1} - \cs^{2}\,\,e_{1}{}^{1}\,\ptl_{x}\alpha
& = & -\,\alpha\,\udot_{1} - (\alpha+2\,\beta)\,[\,\cs^{-2}\,
\frac{d^{2}p}{d\mu^{2}}\,(\mu+p)-\cs^{2}\,]\,\udot_{1}
\nonumber \\ 
& & \hsp5 - \ \cs^{2}\,[\,2\,a_{1}\,(\alpha-\beta)
+ 6\,(n_{-}\,\sig_{\times}-n_{\times}\,\sig_{-})\,] \\
\l{sigmdot}
M^{-1}\,\ptl_{t}\sig_{-} + e_{1}{}^{1}\,\ptl_{x}n_{\times}
& = & -\,(\alpha+2\,\beta)\,\sig_{-}
+ 2\sqrt{3}\,\sig_{\times}^{2}
- \sqrt{3}\,\sig_{2}^{2}
- 2\sqrt{3}\,\,n_{-}^{2}
- (\udot_{1}-2\,a_{1})\,n_{\times} \\
\l{ncdot}
M^{-1}\,\ptl_{t}n_{\times} + e_{1}{}^{1}\,\ptl_{x}\sig_{-}
& = & -\,\alpha\,n_{\times} - \udot_{1}\,\sig_{-} \\
\l{sigcdot}
M^{-1}\,\ptl_{t}\sig_{\times} - e_{1}{}^{1}\,\ptl_{x}n_{-}
& = & -\,(\alpha+2\,\beta+2\sqrt{3}\,\sig_{-})\,
\sig_{\times} + (\udot_{1}-2\,a_{1}-2\sqrt{3}\,n_{\times})\,
n_{-} \\
\l{nmdot}
M^{-1}\,\ptl_{t}n_{-} - e_{1}{}^{1}\,\ptl_{x}\sig_{\times}
& = & -\,(\alpha-2\sqrt{3}\,\sig_{-})\,n_{-}
+ (\udot_{1}+2\sqrt{3}\,n_{\times})\,\sig_{\times} \ .
\eea
The characteristic condition for the set (\r{e11dot}) --
(\r{nmdot}) is invariantly given by $0 = Q =
(u^{a}\zeta_{a})^{5}\,[\,(-\,u^{b}\,u^{c}+\cs^{2}\,h^{bc})\,
\zeta_{b}\,\zeta_{c}\,]\,[\,(-\,u^{d}\,u^{e}+1^{2}\,h^{de})\,
\zeta_{d}\,\zeta_{e}\,]^{2}$,  where $\zeta_{a} := \nabla_{a}\phi$
are the past-directed normals to characteristic 3-surfaces ${\cal
C}$:$\{\phi(x^{\mu})=\mbox{const}\}$. The {\em connection\/}
characteristic eigenfields associated with {\em non-zero\/}
characteristic velocities are $\{\,(\alpha\pm\cs\,\udot_{1})/
(1+\cs^{2})^{1/2}\,\}$ with $v = \pm\,\cs$ and $\{\,(\sig_{-}\mp
n_{\times})/\sqrt{2}, \,(\sig_{\times}\pm n_{-})/\sqrt{2}\,\}$ with 
$v = \pm\,1$.

Initial data which is invariant under the transformations of an
Abelian $G_{2}$ isometry group has to satisfy the following set of
constraint equations: \enl

%\pagebreak
\noindent
{\em Initial value constraint equations\/}: \nopagebreak
\bea
\l{g2clapse}
e_{1}{}^{1}\,M^{-1}\,\ptl_{x}M
& = & \udot_{1} \\
\l{g2ccom1}
0 & = & (C_{\rm com})_{1} \ := \ e_{1}{}^{1}\,\ptl_{x}e_{2}{}^{2}
- (a_{1}+\sqrt{3}\,n_{\times})\,e_{2}{}^{2} \\
\l{g2ccom2}
0 & = & (C_{\rm com})_{2} \ := \ e_{1}{}^{1}\,\ptl_{x}e_{3}{}^{2}
- (a_{1}-\sqrt{3}\,n_{\times})\,e_{3}{}^{2}
+ 2\sqrt{3}\,n_{-}\,e_{2}{}^{2} \\
\l{g2ccom3}
0 & = & (C_{\rm com})_{3} \ := \ e_{1}{}^{1}\,\ptl_{x}e_{3}{}^{3}
- (a_{1}-\sqrt{3}\,n_{\times})\,e_{3}{}^{3} \\
\l{g2cgauss}
0 & = & (C_{\rm G}) \ := \ {}^{*}\!R
+ 2\,(2\,\alpha+\beta)\,\beta
- 6\,(\sig_{-}^{2}+\sig_{\times}^{2}+\sig_{2}^{2})
- 2\,\mu - 2\,\Lambda \\
\l{g2c11}
0 & = & (C_{1})_{1} \ := \ e_{1}{}^{1}\,\ptl_{x}\beta
+ a_{1}\,(\alpha-\beta)
+ 3\,(n_{-}\,\sig_{\times}-n_{\times}\,\sig_{-}) \\
\l{g2c13}
0 & = & (C_{1})_{3} \ := \ (e_{1}{}^{1}\,\ptl_{x}-3\,a_{1}
+\sqrt{3}\,n_{\times})\,\sig_{2} \\
\l{g2cpf}
0 & = & (C_{\rm PF})_{1} \ := \ \cs^{2}\,e_{1}{}^{1}\,\ptl_{x}\mu
+ (\mu+p)\,\udot_{1} \ ;
\eea
the 3-Ricci curvature scalar ${}^{*}\!R$ of spacelike 3-surfaces
${\cal S}$:$\{t=\mbox{const}\}$ orthogonal to ${\bf u}$ is given
by
\be
{}^{*}\!R := 2\,(2\,e_{1}{}^{1}\,\ptl_{x}-3\,a_{1})\,a_{1}
- 6\,(n_{-}^{2}+n_{\times}^{2}) \ .
\ee
The constraint equations (\r{g2cgauss}) -- (\r{g2cpf}) in this set
are specialisations of Eqs. (\r{cfried}), (\r{c11}), (\r{c13jac2})
and (\r{bimom1}), respectively. The remaining ones,
Eqs. (\r{g2clapse}) -- (\r{g2ccom3}), derive from the commutators.
\begin{quotation}
{\em Algorithm\/}: Given an equation of state of the form $p =
p(\mu)$, the {\em initial data\/}, which can be specified {\em
freely\/} (modulo minimal differentiability requirements and the
remaining coordinate reparameterisation freedom) as functions of
the spatial coordinate $x$ on a spacelike 3-surface ${\cal
S}$:$\{t=\mbox{const}\}$ orthogonal to the fluid 4-velocity field
${\bf u}$, are the values of the variables $\{\,e_{1}{}^{1},
\,\alpha, \,\sig_{-}, \,n_{\times}, \,\sig_{\times}, \,n_{-},
\,\mu\,\}$, together with the cosmological constant
$\Lambda$. Specifying the values of the variables $\{\,e_{2}{}^{2},
\,e_{3}{}^{2}, \,e_{3}{}^{3}, \,\beta,\, \sig_{2}, \,a_{1}\,\}$
{\em at one point\/} on this 3-surface, their spatial distribution
follows from Eqs. (\r{g2ccom1}) -- (\r{g2cgauss}), respectively,
while $\udot_{1}$ is determined through Eq. (\r{g2cpf}) and $M$
through Eq. (\r{g2clapse}). Then all {\em time derivatives\/} of
these variables are known. Note that one of the coordinate
components $e_{1}{}^{2}$ and $e_{1}{}^{3}$ is arbitrary as
functions of $x$, too, while the other follows from the
unit-magnitude property of $\p_{1}$. It remains to specify {\em
boundary conditions\/} for all variables to obtain unique
solutions.
\end{quotation}
Two subcases of importance are contained within the class of
Abelian $G_{2}$ perfect fluid cosmological models: the {\em
orthogonally transitive\/} subcase arises when $0 = e_{1}{}^{2} =
e_{1}{}^{3} \Rightarrow 0 = \sig_{2}$, which itself specialises to
the {\em diagonal\/} (``polarised'') subcase when additionally $0 =
e_{3}{}^{2} \Rightarrow 0 = \sig_{\times} = n_{-}$, leading to a
diagonal line element (see Refs. \ct{hewwai90} and \ct{waiell97}).

The initial value constraint equations (\r{g2ccom1}) -- (\r{g2cpf})
are propagated along ${\bf u}$ via a FOSH evolution system of
autonomous partial differential equations, where the characteristic
speeds are $|\,v\,| = 0$, according to \enl

\noindent
{\em 7-dimensional autonomous constraint evolution system\/}:
\nopagebreak
\bea
M^{-1}\,\ptl_{t}(C_{\rm com})_{1}
& = & -\,(\alpha+\beta+\sqrt{3}\,\sig_{-})\,
(C_{\rm com})_{1} - e_{2}{}^{2}\,(C_{1})_{1} \\
M^{-1}\,\ptl_{t}(C_{\rm com})_{2}
& = & -\,(\alpha+\beta-\sqrt{3}\,\sig_{-})\,
(C_{\rm com})_{2} - 2\sqrt{3}\,\sig_{\times}\,
(C_{\rm com})_{1} - e_{3}{}^{2}\,(C_{1})_{1} \\
M^{-1}\,\ptl_{t}(C_{\rm com})_{3}
& = & -\,(\alpha+\beta-\sqrt{3}\,\sig_{-})\,
(C_{\rm com})_{3} - e_{3}{}^{3}\,(C_{1})_{1} \\
M^{-1}\,\ptl_{t}(C_{\rm G})
& = & -\,(\alpha+\beta)\,(C_{\rm G})
- 4\,(\udot_{1}+a_{1})\,(C_{1})_{1} \\
M^{-1}\,\ptl_{t}(C_{1})_{1}
& = & -\,(\alpha+3\,\beta)\,(C_{1})_{1} + \sig_{2}\,
(C_{1})_{3} - \sfrac{1}{4}\,(\udot_{1}-a_{1})\,(C_{\rm G})
- \sfrac{1}{2}\,(C_{\rm PF})_{1} \\
M^{-1}\,\ptl_{t}(C_{1})_{3}
& = & -\,(\alpha+3\,\beta-\sqrt{3}\,\sig_{-})\,(C_{1})_{3}
- 9\,\sig_{2}\,(C_{1})_{1} \\
M^{-1}\,\ptl_{t}(C_{\rm PF})_{1}
& = & -\,2\,(\alpha+\beta)\,(C_{\rm PF})_{1}
- (\alpha+2\,\beta)\,[\,\cs^{2}
+ \cs^{-2}\,\frac{d^{2}p}{d\mu^{2}}\,(\mu+p)\,]\,(C_{\rm PF})_{1}
- 2\,\cs^{-2}\,(\mu+p)\,(C_{1})_{1} \ .
\eea
The full set of propagation equations for the Abelian $G_{2}$
perfect fluid cosmological models now forms a larger autonomous
FOSH evolution system according to Eq. (\r{fosh}), with the
previous 11-dimensional system as a subset. The virtue of the
larger system is that it explicitly shows that the set of dynamical
field equations introduced is consistent: the initial value
constraint equations are preserved by the time evolution equations
(if they are true initially, they remain true thereafter). This
completes the discussion on a well-posed initial value problem for
this class of cosmological models in the $1+3$ orthonormal frame
{\em connection\/} formulation of gravitational fields. \enl

We finally list the specialisations which the expressions
(\r{elep}) -- (\r{tweyl2}) for the Weyl curvature characteristic
eigenfields undergo by the geometrical restrictions imposed by the
Abelian $G_{2}$ isometry group. We use Eq. (\r{g2c13}) to eliminate
derivatives $\ptl_{x}\sigtw$. \enl

\noindent
{\em Weyl curvature characteristic eigenfields\/}:
\bea
\l{cweyle}
E_{+}
& = & -\,\sfrac{1}{3}\,e_{1}{}^{1}\,\ptl_{x}a_{1}
- \sfrac{1}{3}\,(\alpha-\beta)\,\beta
- \sigm^{2} - \sigc^{2} + 2\,\nm^{2} + 2\,\nc^{2}
+ \sfrac{1}{2}\,\sigtw^{2} \\
\l{cweylh}
H_{+}
& = & -\,\sfrac{3}{2}\,(\sigm-\nc)\,(\sigc+\nm)
+ \sfrac{3}{2}\,(\sigm+\nc)\,(\sigc-\nm) \\
\l{g2lweyl1}
(E_{3}\mp H_{2})
& = & -\,\sqrt{3}\,(\sig_{\times}\pm n_{-})\,\sig_{2} \\ 
\l{g2lweyl2}
(E_{2}\pm H_{3})
& = & (\beta\mp a_{1}+\sqrt{3}\,\sig_{-}
\mp \sqrt{3}\,n_{\times})\,\sig_{2} \\
\l{g2tweyl1}
(E_{-}\mp H_{\times})
& = & \mp\,(e_{1}{}^{1}\,\ptl_{x}\mp\alpha-a_{1})\,
(\sig_{-}\mp n_{\times}) \pm 2\sqrt{3}\,n_{-}\,
(\sig_{\times}\pm n_{-}) \pm n_{\times}\,(\beta\mp a_{1})
+ \sfrac{\sqrt{3}}{2}\,\sig_{2}^{2} \\
\l{g2tweyl2}
(E_{\times}\pm H_{-})
& = & \mp\,(e_{1}{}^{1}\,\ptl_{x}\mp\alpha-a_{1})\,
(\sig_{\times}\pm n_{-}) \mp 2\sqrt{3}\,n_{-}\,
(\sig_{-}\mp n_{\times}) \mp n_{-}\,(\beta\mp a_{1}) \ .
\eea
Note that $0 = \sigtw \Rightarrow 0 = (E_{3}\mp H_{2}) = (E_{2}\pm
H_{3})$ holds.

%------------------------------------------------------------------
\subsection{Transport equations for jump discontinuities in outward
first derivatives}
%------------------------------------------------------------------
To give a graphic example, which, we believe, will also be of some
interest in numerical investigations of dynamical features of
Abelian $G_{2}$ perfect fluid cosmological models, we conclude this
section with a brief derivation of the transport equations that
describe how physically relevant jump discontinuities in the
outward first derivatives of the initial data are propagated along
the bicharacteristic rays of the setting. We confine ourselves to
modes with $v \neq 0$ relative to ${\bf u}$. To this end, we first
turn to Eq. (\r{jumpisr}) for each of $v \in \{\,\pm\,\cs,
\,\pm\,1\,\}$, leading to the conditions
\be
\l{g2jumpcon}
[\ptl_{\phi}\udot_{1}] = \pm\,\cs\,[\ptl_{\phi}\alpha] \ ,
\hsp5
[\ptl_{\phi}\nc] = \mp\,[\ptl_{\phi}\sigm] \ ,
\hsp5
[\ptl_{\phi}\nm] = \pm\,[\ptl_{\phi}\sigc] \ ,
\ee
respectively. This then gives from Eq. (\r{jumptrans}), together
with the evolution subsystem (\r{alphadot}) -- (\r{nmdot}), the
relations:
\enl

\noindent
(i) {\em $v = \pm\,\cs$ longitudinal modes\/}: \nopagebreak
\bea
0 & = & (\,M^{-1}\,\ptl_{t} \mp \cs\,e_{1}{}^{1}\,\ptl_{x}
+ \sfrac{1}{2}\,\cs^{-1}\,(M^{-1}\,\ptl_{t}\cs\mp\cs\,e_{1}{}^{1}
\,\ptl_{x}\cs)
\nonumber \\
& & \hspace{15mm} + \ \alpha
+ \sfrac{1}{2}\,(\alpha+2\beta)\,[\,\cs^{-2}\,
\frac{d^{2}p}{d\mu^{2}}\,(\mu+p)-\cs^{2}\,]
\mp \sfrac{1}{2}\,\cs\,\udot_{1} \pm \cs\,a_{1}\,)\,
[\ptl_{\phi}(\alpha\pm\cs^{-1}\,\udot_{1})] \ ;
\eea
jump discontinuities in the outward first derivatives of initial
data for $\{\,\alpha, \,\udot_{1}\,\}$, subject to
Eq. (\r{g2jumpcon}), travel along the local sound cones. Note that,
because of the general functional dependence $\cs =
\cs(\mu)$, the evolution of sound cone initial data typically leads
to the formation of ``shocks''. This phenomenon becomes impossible
in the special case of a {\em linear\/} baryotropic equation of
state with $p(\mu) = (\gam-1)\,\mu$, $1 \leq \gam \leq 2$, where
$\cs = (\gam-1)^{1/2} = \mbox{const}$. \enl

\noindent
(ii) {\em $v = \pm\,1$ transverse modes\/}: \nopagebreak
\bea
0 & = & (M^{-1}\,\ptl_{t}\mp e_{1}{}^{1}\,\ptl_{x}+\alpha+\beta
\mp\udot_{1}\pm a_{1})\,[\ptl_{\phi}(\sigm\mp\nc)] \\
0 & = & (M^{-1}\,\ptl_{t}\mp e_{1}{}^{1}\,\ptl_{x}+\alpha+\beta
\mp\udot_{1}\pm a_{1})\,[\ptl_{\phi}(\sigc\pm\nm)]\ .
\eea
jump discontinuities in the outward first derivatives of initial
data for $\{\,\sigm, \,\nc\,\}$ and $\{\,\sigc, \,\nm\,\}$, subject
to Eq. (\r{g2jumpcon}), travel along the local light cones.

%%%%%%%%%%%%%%%%%%%%%%%%%%%%%%%%%%%%%%%%%%%%%%%%%%%%%%%%%%%%%%%%%%%
\section{Conclusion}
\l{sec:conc}
%%%%%%%%%%%%%%%%%%%%%%%%%%%%%%%%%%%%%%%%%%%%%%%%%%%%%%%%%%%%%%%%%%%
The main conclusion of this work is that in examining a set of
dynamical equations for a physical system such as the relativistic
gravitational field equations, completed by the equations for all
needed auxiliary variables, the constraint equations are crucial in
determining what information can be propagated along the
characteristic 3-surfaces of the evolution equations when these are
expressed in FOSH format. We have explicitly shown how to examine
the constraint equations to determine whether jump discontinuities
in the derivatives of the initial data can be propagated along the
various characteristic 3-surfaces in the case of the relativistic
gravitational field equations with a baryotropic perfect fluid
matter source, including the Weyl curvature variables. This makes
it clear that such an investigation is needed to complement the
determination of the set of characteristic 3-surfaces of any FOSH
evolution system with any existing supplementary constraint
equations, in order to determine which characteristic 3-surfaces in
the set are physically relevant.

The process outlined enables us to show why the characteristic
3-surfaces for the semi-longitudinal Weyl curvature characteristic
eigenfields $(E_{3}\mp H_{2})$ and $(E_{2}\pm H_{3})$ that are
associated with $|\,v\,| = \sfrac{1}{2}$, apparent in a
straightforward reduction of order $\ptl^{3}g$ of the evolution
system of the relativistic gravitational field equations for a
baryotropic perfect fluid matter source to FOSH format, cannot in
fact be activated. It demonstrates that potentially associated
semi-longitudinal gravitational radiation cannot occur, despite the
occurrence of related characteristic 3-surfaces in the FOSH
evolution system. It should be noted that the issue is {\em not\/}
that the constraint equations are incompatible with the evolution
equations, in the sense of not being conserved under the system's
time evolution. On the contrary, the propagation of these
constraint equations is indeed compatible with the existence of
these modes, as can be shown by considering an extended FOSH
evolution system that includes variables representing satisfaction
of the constraint equations \ct{maa97,friren00} (see also
Refs. \ct{vel97}, \ct{hve98} and \ct{mac98}). The issue is that the
constraint equations do not allow the setting of jump
discontinuities in (the derivatives of) the initial data for
$(E_{3}\mp H_{2})$ and $(E_{2}\pm H_{3})$ {\em because the values
of the components of the constraint equations themselves cannot
suffer jump discontinuities\/}: they have to be continuously zero
from one spacetime event to any nearby one, i.e., everywhere.

What this analysis does not do is to show in what manner the
semi-longitudinal Weyl curvature characteristic eigenfields
$(E_{3}\mp H_{2})$ and $(E_{2}\pm H_{3})$, originally identified by
Szekeres and shown there to have observable physical effects
\ct{sze65}, will evolve in time, nor does it adequately
characterise what freedom there is in setting initial data for
these modes. It would be helpful to have some characterisation of
the full freedom to assign these modes on an initial data
3-surface.

The worked example for perfect fluid cosmological models with an
Abelian $G_{2}$ isometry group presented in Sec. \r{sec:g2fosh}
features neatly the conceptual and mathematical advantages one can
gain from combining the idea of FOSH evolution systems with a $1+3$
orthonormal frame {\em connection\/} formulation of gravitational
fields. It will be usable as a multi-facet test bed for numerical
experiments of spacetime geometry evolution processes in
relativistic cosmology and already provides the basis for
work-in-progress on an interesting new scale-invariant,
dimensionless dynamical formulation for the orthogonally transitive
Abelian $G_{2}$ perfect fluid cosmological models \ct{uggetal00}.

%%%%%%%%%%%%%%%%%%%%%%%%%%%%%%%%%%%%%%%%%%%%%%%%%%%%%%%%%%%%%%%%%%%
\acknowledgments
%%%%%%%%%%%%%%%%%%%%%%%%%%%%%%%%%%%%%%%%%%%%%%%%%%%%%%%%%%%%%%%%%%%
We are grateful to J\"{u}rgen Ehlers and Claes Uggla for very
helpful comments. This work was supported by the Deutsche
Forschungsgemeinschaft (DFG) at Bonn, Germany (HvE), and the
Foundation for Research and Development (FRD) at Pretoria, South
Africa (GFRE). In parts the computer algebra packages {\tt REDUCE}
and {\tt CLASSI} were employed.

%%%%%%%%%%%%%%%%%%%%%%%%%%%%%%%%%%%%%%%%%%%%%%%%%%%%%%%%%%%%%%%%%%%
\appendix
\section*{}
%%%%%%%%%%%%%%%%%%%%%%%%%%%%%%%%%%%%%%%%%%%%%%%%%%%%%%%%%%%%%%%%%%%
%------------------------------------------------------------------
\subsection{Suggestive example: linearised relativistic
gravitational field equations in a metric approach with
Regge--Wheeler coordinate gauge fixing}
\l{app1}
%------------------------------------------------------------------
Rooted in tradition, a {\em metric approach\/} to formulating
relativistic gravitational dynamics is still more frequently
encountered in the literature than, e.g., the (extended) $1+3$
orthonormal frame formulation employed in this paper and in
Ref. \ct{hveell99}. This motivates the brief discussion of a metric
approach based example derived from a paper by Kind, Ehlers and
Schmidt \ct{kinetal93} that highlights the issue of the physical
role of the constraint equations in the selection of appropriate
initial data sets. The example arises as a special subcase of
linearised relativistic gravitational field equations that describe
small adiabatic, non-radial perturbations of a star in hydrostatic
equilibrium. Starting from a set of local coordinates $\{\,t, \,r,
\,\vartheta, \,\varphi\,\}$ and imposing Regge--Wheeler coordinate
gauge fixing conditions, the Ansatz for the line element contains
two real-valued metric functions $f = f(t,r)$ and $g = g(t,r)$; the
angular behaviour of the line element can be given in terms of
spherical harmonic functions $Y_{lm}(\vartheta,\varphi)$ (for more
details see Sec. 2 of Ref. \ct{kinetal93}). One can then obtain
from the linearised Einstein field equations a {\em second-order\/}
symmetric hyperbolic evolution system for $f$ and $g$ that takes
the form
\bea
\l{app1eq1}
-\,\ddot{f} + f^{\prime\prime}
& = & \frac{2}{r}\,f^{\prime} - \frac{4}{r^{2}}\,g^{\prime}
+ \frac{l(l+1)}{r^{2}}\,f \\
\l{app1eq2}
-\,\ddot{g} + g^{\prime\prime}
& = & -\,\frac{2}{r}\,g^{\prime} + \frac{2}{r^{2}}\,f
+ \frac{1}{r^{2}}\,(\,l(l+1)-2\,)\,g \ .
\eea
Moreover, $f$ and $g$ are bound to satisfy the constraint equation
\be
\l{app1eq3}
0 = (C) := g^{\prime\prime} - \frac{1}{r}\,f^{\prime}
+ \frac{3}{r}\,g^{\prime} - \frac{l(l+1)}{2r^{2}}\,(f+g)
- \frac{1}{r}\,(f-g) \ .
\ee
The characteristic 3-surfaces ${\cal C}$ determined by the
principal parts of Eqs. (\r{app1eq1}) and (\r{app1eq2}) satisfy the
conditions $(t\mp r) = \mbox{const}$. If we assume that
Eqs. (\r{app1eq1}) and (\r{app1eq2}) hold for $f$ and $g$, then we
obtain for $(C)$ the evolution equation
\be
\l{app1eq4}
0 = -\,(\ddot{C}) + (C)^{\prime\prime} + \frac{2}{r}\,(C)^{\prime}
- \frac{l(l+1)}{2r^{2}}\,(C) \ .
\ee
Hence, if on a spacelike 3-surface ${\cal S}$:$\{t=\mbox{const}\}$
real-valued initial data $I_{\ptl^{2}g} = \{\,f, \,\dot{f}, \,g,
\,\dot{g}\,\}$ satisfy $0 = (C) = (\dot{C})$, then any solution of
Eqs. (\r{app1eq1}) and (\r{app1eq2}) satisfies $0 = (C)$
everywhere.

It is instructive to note that from Eq. (\r{app1eq3}) we can now
add any arbitrary real-valued multiple $a\,g^{\prime\prime}$ to
Eq. (\r{app1eq2}) to obtain a principal part of the form
$$
-\,\ddot{g} + (1+a)\,g^{\prime\prime} \ ,
$$
which instead defines characteristic 3-surfaces ${\cal C}$ that
satisfy the condition $(\,(1+a)^{1/2}\,t\mp r\,) = \mbox{const}$.
In particular, we can choose $a = -\,1$, leading to $r =
\mbox{const}$, which corresponds to a propagation speed $|\,v\,| =
0$. This implies that no information is transported from one event
to a spatially separated nearby one. A different view considers the
differentiability properties of $f$ and $g$: if one chooses
real-valued initial data such that $\{\,f, \,f^{\prime}, \,g,
\,g^{\prime}\,\}$ are {\em continuous\/}, then the constraint
equation (\r{app1eq3}) imposes {\em no condition\/} on the
differentiability of $f^{\prime\prime}$ (\,and so on $\ddot{f}$,
from Eq. (\r{app1eq1})\,), while $g^{\prime\prime}$ (\,and so
$\ddot{g}$, from Eq. (\r{app1eq2})\,) is required to be {\em
continuous\/} too. Hence, jump discontinuities are physically
allowed only in the initial value of the former quantity and are
propagated via Eq. (\r{app1eq1}).

%------------------------------------------------------------------
\subsection{Constraint equations component-wise}
\l{app2}
%------------------------------------------------------------------

\noindent
{\em Fluid shear divergence equations/Jacobi constraint
equations\/}: \nopagebreak
\bea
\l{c11}
0 & = & -\,\sfrac{1}{2}\,(C_{1})_{1}
= \p_{1}(\sfrac{1}{3}\,\Th+\sigp)
- \sfrac{\sqrt{3}}{2}\,(\p_{2}-3\,a_{2}+\sqrt{3}\,\ntw)\,(\sigth)
- \sfrac{\sqrt{3}}{2}\,(\p_{3}-3\,a_{3}-\sqrt{3}\,\nth)\,(\sigtw)
\nonumber \\
& & \hsp5 - \ \sfrac{1}{2}\,(\p_{2}+2\,\udot_{2}-a_{2}
-\sqrt{3}\,\ntw)\,(\om_{3})
+ \sfrac{1}{2}\,(\p_{3}+2\,\udot_{3}-a_{3}+\sqrt{3}\,\nth)\,(\om_{2})
\nonumber \\
& & \hsp5 - \ 3\,a_{1}\,\sigp + 3\,(\nm\,\sigc-\nc\,\sigm)
+ \sfrac{1}{6}\,(n-2\,\np)\,\om_{1}
\\ \nonumber \\
\l{cjac1}
0 & = & (C_{\rm J})_{1}
= \sfrac{1}{3}\,(\p_{1}-2\,a_{1})\,(n-2\np)
+ (\p_{2}-2\,a_{2})\,(a_{3}+\sqrt{3}\nth)
+ (\p_{3}-2\,a_{3})\,(a_{2}-\sqrt{3}\ntw) \nonumber \\
& & \hsp5 + \ 2\,(\sfrac{1}{3}\,\Th-2\,\sigp)\,\om_{1}
+ 2\,(\sqrt{3}\sigtw+\Om_{2})\,\om_{3}
+ 2\,(\sqrt{3}\sigth-\Om_{3})\,\om_{2}
\\ \nonumber \\
\l{c12jac3}
0 & = & \sfrac{1}{\sqrt{3}}\,[\,(C_{1})_{2}\mp (C_{\rm J})_{3}\,]
= \p_{1}(\sigth\mp\ntw-\sfrac{1}{\sqrt{3}}\,\om_{3}
\mp\sfrac{1}{\sqrt{3}}\,a_{2})
+ \p_{2}(\sigm\mp\nc\pm\sfrac{1}{\sqrt{3}}\,a_{1})
+ \p_{3}(\sigc\pm\nm+\sfrac{1}{\sqrt{3}}\,\om_{1}) \nonumber \\
& & \hsp5 + \ \sfrac{1}{\sqrt{3}}\,(\p_{2}-3\,a_{2}
+3\sqrt{3}\,\ntw)\,(\sigp)
\mp \sfrac{1}{3\sqrt{3}}\,(\p_{3}-2\,a_{3})\,(n+\np) 
- \sfrac{2}{3\sqrt{3}}\,\p_{2}(\Th) \nonumber \\
& & \hsp5 - \ 3\,a_{1}\,(\sigth\mp\ntw)
- 3\,a_{2}\,(\sigm\mp\nc) - 3\,a_{3}\,(\sigc\pm\nm)
- (\np-\sqrt{3}\,\nm)\,\sigtw \nonumber \\
& & \hsp5 - \ \nc\,(\sqrt{3}\,\sigth\pm a_{2})
- \ntw\,(\sqrt{3}\,\sigm\pm a_{1}) + \sqrt{3}\,\nth\,\sigc
\pm a_{3}\,\nm \nonumber \\
& & \hsp5 \mp\ \sfrac{1}{\sqrt{3}}\,(2\sqrt{3}\,\sigtw
\mp 2\,\udot_{3}-2\,\Om_{2}\pm a_{3}\pm\sqrt{3}\,\nth)\,\om_{1}
\mp \sfrac{1}{\sqrt{3}}\,(2\sqrt{3}\,\sigc+2\,\Om_{1}
\pm\sfrac{1}{3}\,n\pm\sfrac{1}{3}\,\np\pm\sqrt{3}\,\nm)\,\om_{2}
 \nonumber \\
& & \hsp5 \mp\ \sfrac{1}{\sqrt{3}}\,(\sfrac{2}{3}\,\Th+2\,\sigp
-2\sqrt{3}\,\sigm\pm 2\,\udot_{1}\mp a_{1}\pm\sqrt{3}\,\nc)\,\om_{3}
\\ \nonumber \\
\l{c13jac2}
0 & = & \sfrac{1}{\sqrt{3}}\,[\,(C_{1})_{3}\pm (C_{\rm J})_{2}\,]
= \p_{1}(\sigtw\pm\nth+\sfrac{1}{\sqrt{3}}\,\om_{2}
\mp\sfrac{1}{\sqrt{3}}\,a_{3})
+ \p_{2}(\sigc\pm\nm-\sfrac{1}{\sqrt{3}}\,\om_{1})
- \p_{3}(\sigm\mp\nc\mp\sfrac{1}{\sqrt{3}}\,a_{1}) \nonumber \\
& & \hsp5 \pm\ \sfrac{1}{3\sqrt{3}}\,(\p_{2}-2\,a_{2})\,(n+\np) 
+ \sfrac{1}{\sqrt{3}}\,(\p_{3}-3\,a_{3}-3\sqrt{3}\,\nth)\,(\sigp)
- \sfrac{2}{3\sqrt{3}}\,\p_{3}(\Th) \nonumber \\
& & \hsp5 - \ 3\,a_{1}\,(\sigtw\pm\nth)
- 3\,a_{2}\,(\sigc\pm\nm) + 3\,a_{3}\,(\sigm\mp\nc)
+ (\np+\sqrt{3}\,\nm)\,\sigth \nonumber \\
& & \hsp5 + \ \nc\,(\sqrt{3}\,\sigtw\pm a_{3})
- \nth\,(\sqrt{3}\,\sigm\mp a_{1}) - \sqrt{3}\,\ntw\,\sigc
\pm a_{2}\,\nm \nonumber \\
& & \hsp5 \pm\ \sfrac{1}{\sqrt{3}}\,(2\sqrt{3}\,\sigth
\mp 2\,\udot_{2}+2\,\Om_{3}\pm a_{2}\mp\sqrt{3}\,\ntw)\,\om_{1}
\pm \sfrac{1}{\sqrt{3}}\,(\sfrac{2}{3}\,\Th+2\,\sigp
+2\sqrt{3}\,\sigm\pm 2\,\udot_{1}\mp a_{1}\mp\sqrt{3}\,\nc)\,\om_{2}
\nonumber \\
& & \hsp5 \pm\ \sfrac{1}{\sqrt{3}}\,(2\sqrt{3}\,\sigc-2\,\Om_{1}
\mp\sfrac{1}{3}\,n\mp\sfrac{1}{3}\,\np\pm\sqrt{3}\,\nm)\,\om_{3}
\ .
\eea

\noindent
{\em Fluid vorticity divergence equation\/}: \nopagebreak
\be
\l{cdivom}
0 = (C_{2}) = (\p_{1}-\udot_{1}-2\,a_{1})\,(\om_{1})
+ (\p_{2}-\udot_{2}-2\,a_{2})\,(\om_{2})
+ (\p_{3}-\udot_{3}-2\,a_{3})\,(\om_{3}) \ .
\ee

\noindent
{\em Generalised Gau\ss--Friedmann equation\/}: \nopagebreak
\bea
\l{cfried}
0 = (C_{\rm G})
& = & 2\,(2\,\p_{1}-3\,a_{1})\,(a_{1})
+ 2\,(2\,\p_{2}-3\,a_{2})\,(a_{2})
+ 2\,(2\,\p_{3}-3\,a_{3})\,(a_{3}) \nonumber \\
& & \hsp5 + \ \sfrac{1}{6}\,n^{2} - \sfrac{2}{3}\,\np^{2}
- 6\,(\nm^{2} + \nc^{2} + \ntw^{2} + \nth^{2}) \nonumber \\
& & \hsp5 + \ 6\,(\sfrac{1}{3}\,\Th-\sigp)\,(\sfrac{1}{3}\,\Th
+\sigp) - 6\,(\sigm^{2} + \sigc^{2} + \sigtw^{2} + \sigth^{2})
 \nonumber \\
& & \hsp5 + \ 2\,(\om_{1}-2\,\Om_{1})\,\om_{1}
+ 2\,(\om_{2}-2\,\Om_{2})\,\om_{2}
+ 2\,(\om_{3}-2\,\Om_{3})\,\om_{3} - 2\,\mu - 2\,\Lambda \ .
\eea

\noindent
{\em Weyl curvature divergence equations\/}: \nopagebreak
\bea
\l{c41}
0 & = & -\,\sfrac{1}{2}\,(C_{4})_{1}
= (\p_{1}-3\,a_{1})\,(E_{+})
- \sfrac{\sqrt{3}}{2}\,(\p_{2}-3\,a_{2}+\sqrt{3}\,\ntw)\,(E_{3})
- \sfrac{\sqrt{3}}{2}\,(\p_{3}-3\,a_{3}-\sqrt{3}\,\nth)\,(E_{2})
\nonumber \\
& & \hsp5 + \ \sfrac{1}{6}\,\p_{1}(\mu) + 3\,\nm\,E_{\times}
- 3\,\nc\,E_{-} + 3\,\sigm\,H_{\times} - 3\,\sigc\,H_{-}
\nonumber \\
& & \hsp5 - \ 3\,\om_{1}\,H_{+}
- \sfrac{3}{2}\,(\sigtw-\sqrt{3}\,\om_{2})\,H_{3}
+ \sfrac{3}{2}\,(\sigth+\sqrt{3}\,\om_{3})\,H_{2}
\\ \nonumber \\
\l{c51}
0 & = & -\,\sfrac{1}{2}\,(C_{5})_{1}
= (\p_{1}-3\,a_{1})\,(H_{+})
- \sfrac{\sqrt{3}}{2}\,(\p_{2}-3\,a_{2}+\sqrt{3}\,\ntw)\,(H_{3})
- \sfrac{\sqrt{3}}{2}\,(\p_{3}-3\,a_{3}-\sqrt{3}\,\nth)\,(H_{2})
\nonumber \\
& & \hsp5 - \ \sfrac{1}{2}\,(\mu+p)\,\om_{1} + 3\,\nm\,H_{\times}
- 3\,\nc\,H_{-} - 3\,\sigm\,E_{\times} + 3\,\sigc\,E_{-}
\nonumber \\
& & \hsp5 + \ 3\,\om_{1}\,E_{+}
+ \sfrac{3}{2}\,(\sigtw-\sqrt{3}\,\om_{2})\,E_{3}
- \sfrac{3}{2}\,(\sigth+\sqrt{3}\,\om_{3})\,E_{2}
\\ \nonumber \\
\l{c42c53}
0 & = & \sfrac{1}{\sqrt{3}}\,[\,(C_{4})_{2}\mp (C_{5})_{3}\,]
= (\p_{1}\pm 3\,\sigp-3\,a_{1})\,(E_{3}\mp H_{2})
+ (\p_{2}\mp\sqrt{3}\,\sigth\pm 3\,\om_{3}-3\,a_{2}
-\sqrt{3}\,\ntw)\,(E_{-}\mp H_{\times}) \nonumber \\
& & \hsp5 + \ (\p_{3}\mp\sqrt{3}\,\sigtw\mp3\,\om_{2}-3\,a_{3}
+\sqrt{3}\,\nth)\,(E_{\times}\pm H_{-})
+ \sfrac{1}{\sqrt{3}}\,(\p_{2}\mp 3\sqrt{3}\,\sigth\mp 3\,\om_{3}
-3\,a_{2}+3\sqrt{3}\,\ntw)\,(E_{+}) \nonumber \\
& & \hsp5 \mp\ \sfrac{1}{\sqrt{3}}\,(\p_{3}\mp 3\sqrt{3}\,\sigtw
\pm 3\,\om_{2}-3\,a_{3}-3\sqrt{3}\,\nth)\,(H_{+})
- \sfrac{1}{3\sqrt{3}}\,\p_{2}(\mu)
\mp \sfrac{1}{\sqrt{3}}\,(\mu+p)\,\om_{3} \nonumber \\
& & \hsp5 \pm\ \sqrt{3}\,(\sigm\mp\nc)\,(E_{3}\pm H_{2})
\pm \sqrt{3}\,(\sigc\pm\nm)\,(E_{2}\mp H_{3})
\mp (3\,\om_{1}\pm\np)\,(E_{2}\pm H_{3})
\\ \nonumber \\
\l{c43c52}
0 & = & \sfrac{1}{\sqrt{3}}\,[\,(C_{4})_{3}\pm (C_{5})_{2}\,]
= (\p_{1}\pm 3\,\sigp-3\,a_{1})\,(E_{2}\pm H_{3})
+ (\p_{2}\mp\sqrt{3}\,\sigth\pm 3\,\om_{3}-3\,a_{2}
-\sqrt{3}\,\ntw)\,(E_{\times}\pm H_{-}) \nonumber \\
& & \hsp5 - \ (\p_{3}\mp\sqrt{3}\,\sigtw\mp3\,\om_{2}-3\,a_{3}
+\sqrt{3}\,\nth)\,(E_{-}\mp H_{\times})
\pm \sfrac{1}{\sqrt{3}}\,(\p_{2}\mp 3\sqrt{3}\,\sigth\mp 3\,\om_{3}
-3\,a_{2}+3\sqrt{3}\,\ntw)\,(H_{+}) \nonumber \\
& & \hsp5 + \ \sfrac{1}{\sqrt{3}}\,(\p_{3}\mp 3\sqrt{3}\,\sigtw
\pm 3\,\om_{2}-3\,a_{3}-3\sqrt{3}\,\nth)\,(E_{+})
- \sfrac{1}{3\sqrt{3}}\,\p_{3}(\mu)
\pm \sfrac{1}{\sqrt{3}}\,(\mu+p)\,\om_{2} \nonumber \\
& & \hsp5 \mp\ \sqrt{3}\,(\sigm\mp\nc)\,(E_{2}\mp H_{3})
\pm \sqrt{3}\,(\sigc\pm\nm)\,(E_{3}\pm H_{2})
\pm (3\,\om_{1}\pm\np)\,(E_{3}\mp H_{2}) \ .
\eea

\noindent
{\em Momentum conservation equations\/}: \nopagebreak
\bea
\l{bimom1}
0 = (C_{\rm PF})_{1}
& = & \cs^{2}\,\p_{1}(\mu) + (\mu+p)\,\udot_{1} \\
\l{bimom2}
0 = (C_{\rm PF})_{2}
& = & \cs^{2}\,\p_{2}(\mu) + (\mu+p)\,\udot_{2} \\
\l{bimom3}
0 = (C_{\rm PF})_{3}
& = & \cs^{2}\,\p_{3}(\mu) + (\mu+p)\,\udot_{3} \ .
\eea

\noindent
{\em Weyl curvature characteristic eigenfields\/}: \nopagebreak
\bea
\l{elep}
E_{+} & = & -\,\sfrac{1}{3}\,\p_{1}(a_{1})
+ \sfrac{1}{6}\,(\p_{2}-3\sqrt{3}\,\ntw)\,(a_{2})
+ \sfrac{1}{6}\,(\p_{3}+3\sqrt{3}\,\nth)\,(a_{3}) \nonumber \\
& & \hsp5 + \ \sfrac{\sqrt{3}}{2}\,(\p_{2}-a_{2}
-\sfrac{1}{\sqrt{3}}\,\ntw)\,(\ntw)
- \sfrac{\sqrt{3}}{2}\,(\p_{3}-a_{3}+\sfrac{1}{\sqrt{3}}\,
\nth)\,(\nth) \nonumber \\
& & \hsp5 + \ (\sfrac{1}{3}\,\Th+\sigp)\,\sigp
+ \sfrac{1}{3}\,(n-2\,\np)\,\np \nonumber \\
& & \hsp5 - \ (\sigm-\nc)\,(\sigm+\nc) - (\sigc+\nm)\,(\sigc-\nm)
\nonumber \\
& & \hsp5 + \ \sfrac{1}{2}\,(\sigth-\ntw)\,(\sigth+\ntw)
+ \sfrac{1}{2}\,(\sigtw+\nth)\,(\sigtw-\nth) \nonumber \\
& & \hsp5 + \ \nm^{2} + \nc^{2}
+ \sfrac{1}{3}\,(\om_{1}-2\,\Om_{1})\,\om_{1}
- \sfrac{1}{6}\,(\om_{2}-2\,\Om_{2})\,\om_{2}
- \sfrac{1}{6}\,(\om_{3}-2\,\Om_{3})\,\om_{3} \nonumber \\
& & \hsp5 + \ \sfrac{1}{2}\,(C_{\rm G})_{11}
\\ \nonumber \\
\l{magp}
H_{+} & = & -\,\sfrac{\sqrt{3}}{2}\,(\p_{2}-a_{2}-
\sqrt{3}\,n_{2})\,(\sigtw)
+ \sfrac{\sqrt{3}}{2}\,(\p_{3}-a_{3}+\sqrt{3}\,n_{3})\,(\sigth)
\nonumber \\
& & \hsp5 + \ \sfrac{1}{3}\,(\p_{1}+2\,\udot_{1}+a_{1})\,(\om_{1})
- \sfrac{1}{6}\,(\p_{2}+2\,\udot_{2}+a_{2}
-3\sqrt{3}\,\ntw)\,(\om_{2}) \nonumber \\
& & \hsp5 - \ \sfrac{1}{6}\,(\p_{3}+2\,\udot_{3}+a_{3}
+3\sqrt{3}\,\nth)\,(\om_{3}) \nonumber \\
& & \hsp5 - \ \sfrac{1}{2}\,(n-2\,\np)\,\sigp
- \sfrac{3}{2}\,(\sigm-\nc)\,(\sigc+\nm)
+ \sfrac{3}{2}\,(\sigm+\nc)\,(\sigc-\nm) \nonumber \\
& & \hsp5 - \ \sfrac{1}{2}\,(C_{3})_{11}
\\ \nonumber \\
\l{lweyl1}
(E_{3}\mp H_{2})
& = & \mp\,\sfrac{1}{2}\,\p_{1}(\sigth\mp\ntw-\sfrac{1}{\sqrt{3}}\, 
\om_{3}\mp\sfrac{1}{\sqrt{3}}\,a_{2})
\pm \sfrac{1}{2}\,\p_{2}(\sigm\mp\nc\pm\sfrac{1}{\sqrt{3}}\,a_{1})
\nonumber \\
& & \hsp5 \pm \ \sfrac{1}{2}\,\p_{3}(\sigc\pm\nm
+\sfrac{1}{\sqrt{3}}\,\om_{1})
\mp\sfrac{\sqrt{3}}{2}\,(\p_{2}-a_{2}+\sqrt{3}\,\ntw)\,(\sigp)
+ \sfrac{1}{2\sqrt{3}}\,(\p_{3}-2\,a_{3})\,(\np) \nonumber \\
& & \hsp5 - \ \sqrt{3}\,(\sig_{2}\pm\sfrac{1}{2\sqrt{3}}\,a_{3}
\mp\sfrac{3}{2}\,\nth)\,(\sigc\pm\nm)
- \sqrt{3}\,(\sig_{3}\pm\sfrac{1}{2\sqrt{3}}\,a_{2}
\pm\sfrac{3}{2}\,\ntw)\,(\sigm\mp\nc) \nonumber \\
& & \hsp5 + \ (\sfrac{1}{3}\,\Th+\sigp)\,\sigth
\pm \sfrac{1}{2}\,a_{1}\,(\sigth\mp 2\,\ntw)
\pm \sfrac{1}{2}\,(n-\np)\,\sigtw
+ \sfrac{1}{3}\,(n-2\,\np)\,\nth \nonumber \\
& & \hsp5 \mp\ \sfrac{\sqrt{3}}{2}\,\nm\,(\sigtw\mp\nth
\pm\sfrac{1}{\sqrt{3}}\,a_{3})
\pm \sfrac{\sqrt{3}}{2}\,\nc\,(\sigth\pm\ntw
\pm\sfrac{1}{\sqrt{3}}\,a_{2}) \nonumber \\
& & \hsp5 \pm\ \sfrac{1}{2\sqrt{3}}\,(2\,\udot_{3}\mp\om_{2}
\pm 2\,\Om_{2}+a_{3}+\sqrt{3}\,\nth)\,\om_{1} \nonumber \\
& & \hsp5 - \ \sfrac{1}{2\sqrt{3}}\,(\om_{1}-2\,\Om_{1}
\mp\np\pm\sqrt{3}\,\nm)\,\om_{2}
\pm \sfrac{1}{2\sqrt{3}}\,(2\,\udot_{1}+a_{1}
-\sqrt{3}\,\nc)\,\om_{3} \nonumber \\
& & \hsp5 - \ \sfrac{1}{\sqrt{3}}\,[\,(C_{\rm G})_{12}
\pm (C_{3})_{31}\,]
\\ \nonumber \\
\l{lweyl2}
(E_{2}\pm H_{3})
& = & \mp\,\sfrac{1}{2}\,\p_{1}(\sigtw\pm\nth+\sfrac{1}{\sqrt{3}}\, 
\om_{2}\mp\sfrac{1}{\sqrt{3}}\,a_{3})
\pm \sfrac{1}{2}\,\p_{2}(\sigc\pm\nm-\sfrac{1}{\sqrt{3}}\,\om_{1})
\nonumber \\
& & \hsp5 \mp\ \sfrac{1}{2}\,\p_{3}(\sigm\mp\nc
\mp\sfrac{1}{\sqrt{3}}\,a_{1})
- \sfrac{1}{2\sqrt{3}}\,(\p_{2}-2\,a_{2})\,(\np)
\mp\sfrac{\sqrt{3}}{2}\,(\p_{3}-a_{3}-\sqrt{3}\,\nth)\,(\sigp)
\nonumber \\
& & \hsp5 + \ \sqrt{3}\,(\sig_{2}\pm\sfrac{1}{2\sqrt{3}}\,a_{3}
\mp\sfrac{3}{2}\,\nth)\,(\sigm\mp\nc)
- \sqrt{3}\,(\sig_{3}\pm\sfrac{1}{2\sqrt{3}}\,a_{2}
\pm\sfrac{3}{2}\,\ntw)\,(\sigc\pm\nm) \nonumber \\
& & \hsp5 + \ (\sfrac{1}{3}\,\Th+\sigp)\,\sigtw
\pm \sfrac{1}{2}\,a_{1}\,(\sigtw\pm 2\,\nth)
\mp \sfrac{1}{2}\,(n-\np)\,\sigth
+ \sfrac{1}{3}\,(n-2\,\np)\,\ntw \nonumber \\
& & \hsp5 \mp\ \sfrac{\sqrt{3}}{2}\,\nm\,(\sigth\pm\ntw
\pm\sfrac{1}{\sqrt{3}}\,a_{2})
\mp \sfrac{\sqrt{3}}{2}\,\nc\,(\sigtw\mp\nth
\pm\sfrac{1}{\sqrt{3}}\,a_{3}) \nonumber \\
& & \hsp5 \mp\ \sfrac{1}{2\sqrt{3}}\,(2\,\udot_{2}\pm\om_{3}
\mp 2\,\Om_{3}+a_{2}-\sqrt{3}\,\ntw)\,\om_{1} \nonumber \\
& & \hsp5 \mp\ \sfrac{1}{2\sqrt{3}}\,(2\,\udot_{1}+a_{1}
+\sqrt{3}\,\nc)\,\om_{2} 
- \sfrac{1}{2\sqrt{3}}\,(\om_{1}-2\,\Om_{1}
\mp\np\mp\sqrt{3}\,\nm)\,\om_{3} \nonumber \\
& & \hsp5 - \ \sfrac{1}{\sqrt{3}}\,[\,(C_{\rm G})_{31}
\mp (C_{3})_{12}\,]
\\ \nonumber \\
\l{tweyl1}
(E_{-}\mp H_{\times})
& = & \mp\,\p_{1}(\sigm\mp\nc)
\pm \sfrac{1}{2}\,\p_{2}(\sigth\mp\ntw+\sfrac{1}{\sqrt{3}}\,\om_{3}
\pm\sfrac{1}{\sqrt{3}}\,a_{2})
\mp \sfrac{1}{2}\,\p_{3}(\sigtw\pm\nth-\sfrac{1}{\sqrt{3}}\,\om_{2}
\pm\sfrac{1}{\sqrt{3}}\,a_{3}) \nonumber \\
& & \hsp5 + \ \sfrac{\sqrt{3}}{2}\,(\sigtw\pm\sfrac{1}{\sqrt{3}}
\,a_{3}\pm 2\,\nth)\,(\sigtw\pm\nth)
- \sfrac{\sqrt{3}}{2}\,(\sigth\pm\sfrac{1}{\sqrt{3}}
\,a_{2}\mp 2\,\ntw)\,(\sigth\mp\ntw) \nonumber \\
& & \hsp5 + \ (\sfrac{1}{3}\,\Th-2\,\sigp)\,\sigm
\pm 3\,\nc\,\sigp \pm \sfrac{1}{2}\,(n+2\,\np)\,\sigc
+ \sfrac{1}{3}\,(n+4\,\np)\,\nm \nonumber \\
& & \hsp5 \pm\ a_{1}\,(\sigm\mp 2\,\nc) + \sfrac{1}{2}\,a_{2}\,\ntw
+ \sfrac{1}{2}\,a_{3}\,\nth \pm \nm\,\om_{1} \nonumber \\
& & \hsp5 \pm\ \sfrac{1}{2\sqrt{3}}\,(2\,\udot_{3}\mp\om_{2}
\pm 2\,\Om_{2}+a_{3}-\sqrt{3}\,\nth)\,\om_{2}
\pm \sfrac{1}{2\sqrt{3}}\,(2\,\udot_{2}\pm\om_{3}
\mp 2\,\Om_{3}+a_{2}+\sqrt{3}\,\ntw)\,\om_{3} \nonumber \\
& & \hsp5 - \ \sfrac{1}{2\sqrt{3}}\,[\,(C_{\rm G})_{22}
- (C_{\rm G})_{33} \pm 2\,(C_{3})_{23}\,]
\\ \nonumber \\
\l{tweyl2}
(E_{\times}\pm H_{-})
& = & \mp\,\p_{1}(\sigc\pm\nm)
\pm \sfrac{1}{2}\,\p_{2}(\sigtw\pm\nth-\sfrac{1}{\sqrt{3}}\,\om_{2}
\pm\sfrac{1}{\sqrt{3}}\,a_{3})
\pm \sfrac{1}{2}\,\p_{3}(\sigth\mp\ntw+\sfrac{1}{\sqrt{3}}\,\om_{3}
\pm\sfrac{1}{\sqrt{3}}\,a_{2}) \nonumber \\
& & \hsp5 - \ \sfrac{\sqrt{3}}{2}\,(\sigtw\pm\sfrac{1}{\sqrt{3}}
\,a_{3}\pm 2\,\nth)\,(\sigth\mp\ntw)
- \sfrac{\sqrt{3}}{2}\,(\sigth\pm\sfrac{1}{\sqrt{3}}
\,a_{2}\mp 2\,\ntw)\,(\sigtw\pm\nth) \nonumber \\
& & \hsp5 + \ (\sfrac{1}{3}\,\Th-2\,\sigp)\,\sigc
\mp 3\,\nm\,\sigp \mp \sfrac{1}{2}\,(n+2\,\np)\,\sigm
+ \sfrac{1}{3}\,(n+4\,\np)\,\nc \nonumber \\
& & \hsp5 \pm\ a_{1}\,(\sigc\pm 2\,\nm) - \sfrac{1}{2}\,a_{2}\,\nth
+ \sfrac{1}{2}\,a_{3}\,\ntw \pm \nc\,\om_{1} \nonumber \\
& & \hsp5 \mp\ \sfrac{1}{2\sqrt{3}}\,(2\,\udot_{2}\pm\om_{3}
\mp 2\,\Om_{3}+a_{2}+\sqrt{3}\,\ntw)\,\om_{2}
\pm \sfrac{1}{2\sqrt{3}}\,(2\,\udot_{3}\mp\om_{2}
\pm 2\,\Om_{2}+a_{3}-\sqrt{3}\,\nth)\,\om_{3} \nonumber \\
& & \hsp5 - \ \sfrac{1}{2\sqrt{3}}\,[\,2\,(C_{\rm G})_{23}
\mp (C_{3})_{22} \pm (C_{3})_{33}\,] \ .
\eea
%

%%%%%%%%%%%%%%%%%%%%%%%%%%%%%%%%%%%%%%%%%%%%%%%%%%%%%%%%%%%%%%%%%%%

%%%%%%%%%%%%%%%%%%%%%%%%%%%%%%%%%%%%%%%%%%%%%%%%%%%%%%%%%%%%%%%%%%%

%\newpage
%
\begin{table}
\caption{Conventions for $(1+1+2)$--decomposition}
\l{tab:112conv}
\begin{tabular}{ll}
{\em Outward\/} frame derivative (across ${\cal J}$): &
$\p_{1}$ \\
{\em Tangential\/} frame derivatives
(along ${\cal J}$): & $\p_{2}$, $\p_{3}$ \\
(Semi-){\em Longitudinal\/} tensor/connection frame components
(wrt. $\p_{1}$): &
$a_{+}$, $a_{2}$, $a_{3}$ \ / \ $n$, $n_{+}$, $n_{2}$, $n_{3}$ \\
{\em Transverse\/} tensor/connection frame components
(wrt. $\p_{1}$): &
$a_{-}$, $a_{\times}$ \ / \ $n_{-}$, $n_{\times}$
\end{tabular}
\end{table}
%

%%%%%%%%%%%%%%%%%%%%%%%%%%%%%%%%%%%%%%%%%%%%%%%%%%%%%%%%%%%%%%%%%%%
\end{document}